\documentclass[conference]{IEEEtran}
\usepackage[utf8]{inputenc}

\usepackage{graphicx}
\usepackage{natbib}
\usepackage{geometry}
\usepackage{graphicx}
\usepackage{fleqn}
\usepackage{hyperref}
\usepackage[bottom]{footmisc}
\graphicspath{{figs/}} 
\usepackage[linesnumbered,ruled,vlined]{algorithm2e}
\usepackage{amsmath,amsthm,amssymb}

\usepackage[pdftex]{color}
\usepackage[font=footnotesize,labelfont=bf]{caption}
\usepackage[usenames,dvipsnames]{xcolor}
\usepackage{ragged2e,booktabs,multirow,tabularx,colortbl}
\usepackage{rotating, makecell}
\usepackage{soul}
\usepackage{tcolorbox}

\newcommand{\etal}{\emph{et al.}}
\usepackage{varwidth}
\usepackage{ragged2e,tikz}
\usetikzlibrary{calc}

\newcounter{tipcounter}
\newcommand{\sys}{\textsf{SCOUT}}

\newcommand{\heuristics}[4][\textwidth-\pgfkeysvalueof{/pgf/inner xsep}-2mm]{%
\begin{center}
\centering
\begin{tikzpicture}
\node[line width=.1mm, rounded corners, draw=#2, inner ysep=5pt, inner xsep=5pt, text width=.95#1, outer sep=0] (one) {\vspace*{13pt}\\\begin{varwidth}{\textwidth}#4\end{varwidth}};
\node[text=white,anchor=north east,align=center, minimum height=1.6em] (two) at (one.north east) {#3};
\path[fill=#2] 
    (one.north west|-two.west) --
    ($(two.west)+(-0.5cm,0)$) 
    to[out=0,in=180] (two.south west) --
    (two.south east) [rounded corners] --
    (one.north east) -- 
    (one.north west) [sharp corners] -- cycle;
\node[text=white,anchor=north east,align=center, minimum height=1em, text height=1.7ex] (three) at (one.north east) {\justifying #3 \hspace*{.5mm}};
\end{tikzpicture} 
\end{center}%
\stepcounter{tipcounter}
}

\begin{document}

\title{A Framework To Improve User Story Sets Through Collaboration}
\author{
\IEEEauthorblockN{Salih Göktuğ Köse}
\IEEEauthorblockA{
Boğaziçi University\\
Istanbul, Turkey \\
salih.kose@boun.edu.tr}\and
\IEEEauthorblockN{Fatma Başak Aydemir}
\IEEEauthorblockA{Boğaziçi University\\
Istanbul, Turkey\\
basak.aydemir@boun.edu.tr}
}
\maketitle
\begin{abstract}
Agile methodologies have become increasingly popular in recent years.  Due to its inherent nature, agile methodologies involve stakeholders with a wide range of expertise and require interaction between them, relying on collaboration and customer involvement. Hence, agile methodologies encourage collaboration between all team members so that more efficient and effective processes are maintained.
Generating requirements can be challenging, as it requires the participation of multiple stakeholders who describe various aspects of the project and possess a shared understanding of essential concepts. One simple method for capturing requirements using natural language is through user stories, which document the agreed-upon properties of a project. Stakeholders try to strive for completeness while generating user stories, but the final user story set may still be flawed.
To address this issue, we propose \sys{}: Supporting Completeness of User Story Sets, which employs a natural language processing pipeline to extract key concepts from user stories and construct a knowledge graph by connecting related terms. The knowledge graph and different heuristics are then utilized to enhance the quality and completeness of the user story sets by generating suggestions for the stakeholders.
We perform a user study to evaluate \sys{} and demonstrate its performance in constructing user stories. The quantitative and qualitative results indicate that \sys{} significantly enhance the quality and completeness of the user story sets.
Our contribution is threefold. First, we develop heuristics to suggest new concepts to include in user stories by considering both the individuals' and other team members' contributions. Second, we implement an open-source collaborative tool to support writing user stories and ensuring their quality. Third, we share the experimental setup and materials. 
\end{abstract}


\section{Introduction} 
\label{section:introduction}


Requirements engineering practices aim to define the various aspects of a software project to ensure that it meets the necessary expectations. In the context of agile methodologies, multiple experts with diverse backgrounds contribute to the formation of viable requirements definitions for different aspects of the project. Collaborative tasks help stakeholders effectively utilize their time and resources \cite{stallinger2001system}. However, aligning the independent activities of these stakeholders can be challenging, particularly when stakeholders are located in different places. While bringing stakeholders together physically may be a solution to this problem, it can be costly in terms of both time and money. This situation may be problematic for projects with strict deadlines and limited resources. Therefore, collaboration is critical in such environments.


Natural language is commonly used for representing requirements documents as it is less time-consuming and resource-intensive compared to other representations such as formal or semi-formal representations \cite{kassab2014state, kassab2015changing}. User stories are a widely-used natural language notation for capturing requirements \cite{lucassen2016use,kassab2015changing}. They follow a predefined format and are therefore utilized by a large number of practitioners in the industry due to their ease of use \cite{dalpiaz2018agile}. During requirements engineering practices, a set of user stories is compiled. Practitioners attempt to ensure that this set covers every detail of the project. However, manually determining the completeness of the user story set can be challenging. In such cases, providing on-demand suggestions to stakeholders when creating user stories can save time and effort that would otherwise be spent examining the user story set. Therefore, an automated system that assists stakeholders in constructing a complete solution can be beneficial.

In collaborative environments, generating requirements is a complex process due to the need for multiple stakeholders to represent various aspects of the project, all of whom must possess a shared understanding of key concepts \cite{gemkow2018automatic}. It is not feasible for stakeholders to continually evaluate the entire set of user stories while introducing different aspects of a system. Additionally, ensuring that all stakeholders are on the same page and providing a coherent solution for a project can be challenging in such environments. To alleviate these difficulties for stakeholders, we propose the use of \sys{} to increase the completeness of user stories in collaborative environments. This system offers an interactive tool that generates on-demand suggestions by constructing a knowledge graph depending on the user input. The knowledge graph is created by employing a natural language processing (NLP) pipeline to extract key concepts from user stories and group them based on their relatedness. We then apply various heuristics to the information stored in knowledge graphs to generate suggestions for stakeholders. Similar to the approach of Aydemir and Dalpiaz~\cite{aydemir2020supporting}, our pipeline relies on natural language text so that incorporating additional features in the future will not require extensive changes. Our pipeline measures the similarity of different noun phrases using sentence embeddings to match slightly different terms that refer to the same concept. We extract keywords using a deep language model and use various algorithms to connect related terms and then construct the knowledge graph. By leveraging the knowledge graph and various heuristics, we generate suggestions for stakeholders to enhance the overall completeness of user stories for a project.

We propose, 
\begin{enumerate}
    \item A web interface with a simplistic design that enables stakeholders to construct requirements artifacts easily.
    \item A NLP pipeline that processes the raw requirements texts. This pipeline extracts concepts from user stories. 
    \item A graph module that constructs knowledge graphs via defining the relationships between these concepts using the state-of-the-art BERT deep language model and storing these relations between concepts.
    \item A suggestion module employs various heuristics to generate suggestions for enhancing the completeness of the user story set, leveraging the knowledge graphs.
    \item A publicly available user story set.\footnote{https://doi.org/10.5281/zenodo.7529130}
    \item A publicly available source code.\footnote{https://doi.org/10.5281/zenodo.7529090} 
\end{enumerate}

The remainder of the paper is structured as follows. Sec. \ref{section:related_work} reviews the related work. Sec. \ref{section:research_approach} explains our research approach. Sec. \ref{section:method} focuses on the details of \sys. Sec. \ref{section:evaluation} reports on the evaluation plan. Finally, Sec. \ref{section:conclusion} concludes the paper.

\section{Related Work} 
\label{section:related_work}

This section presents the selected works from four main related areas: agile requirements engineering, collaboration in software engineering, NLP for requirements engineering, CrowdRE, and gamification.

\subsection{Agile Requirements Engineering}

Kassab \etal{}~\cite{kassab2014state} state that a vast number of practitioners prefer natural language for software requirements representation. Among the free, semi-structured, and structured formats, user stories, which follow a semi-structured notation, are widely preferred by developers adopting agile methodologies~\cite{lucassen2016use,kassab2015changing} due to their simplicity. The popularity of the user stories is the main motivation behind our research.


Despite their simple structure, user stories are not always high quality. Lucassen \etal{} \cite{lucassen2015forging} state that user stories are frequently poor in quality. As with any artifact constructed with natural language, user stories suffer from ambiguity. 


Stakeholders often spare less time to construct requirements than needed. Therefore, defects like ambiguity and incompleteness emerge commonly in software requirements. Dalpiaz \etal{} \cite{dalpiaz2019detecting} propose the REVV-light tool that presents terms in requirements in different viewpoints using Venn diagrams to mainly detect ambiguities and partially detect incompleteness in user stories. Similarly, we focus on a common requirements defect as incompleteness in our research. Additionally, we employ a deep language model to generate suggestions for stakeholders to pinpoint the cause of incompleteness in user stories. Yang \etal{} \cite{yang2011analysing} propose a method to detect anaphoric ambiguity caused by different interpretations of pronouns by different stakeholders via implementing different NLP heuristics. Barbosa \etal{} \cite{barbosa2016use} focus on detecting duplicate user stories that are present in an agile development cycle via using information retrieval metrics along with semantic similarity measures. Yang \etal{} \cite{yang2010automatic} propose a tool that employs a machine learning algorithm to detect coordination ambiguity in requirements texts. Literature shows that defects in natural language requirements are widely studied areas. However, a vast number of studies mainly focus on the intrinsic defects of user stories. Our method focuses on increasing the completeness of user story sets. 

In large-scale software projects, stakeholders often require a better understanding of the requirements definitions. To support the identification of requirements artifacts, conceptual models and goal-oriented approaches are commonly used. Trkman \etal{} \cite{trkman2019impact} experiment with human experts to investigate the effectiveness of conceptual models -specifically business process models- compared to textual requirements and point out that participants achieve equal or better results when a BPMN model is provided. Güneş and Aydemir \cite{gunecs2020automated}, propose a method for capturing relations among user stories via automatically generating goal models using an NLP pipeline to eliminate the human effort required for constructing models. Wautelet \etal{} \cite{wautelet2016building} state that due to its simple structure a large number of user stories are needed in large-scale projects. This leads to several difficulties such as inconsistency. They propose a method that generates a rationale diagram based on i* framework components that allow easier user story analysis. Jaqueira \etal{} \cite{jaqueira2013using} state that user stories have a limited capability for representing and detailing requirements and they propose a method to create a broader viewpoint for stakeholders via enriching user stories with i* framework. Wautelet \etal{} \cite{wautelet2018modelers} experimented with different target groups to investigate the difficulties that modelers encountered while generating goal-oriented models from user story sets. Throughout their experiment, the authors opted for completeness as an evaluation criterion. 

Literature states that user stories are widely used in agile development processes. Even though user stories follow a simple structure, agile practitioners need support when dealing with user stories in large-scale environments to prevent defects such as incompleteness and ambiguity. The basis of our research depends on supporting stakeholders when constructing user stories in terms of completeness.

\subsection{Collaboration in Software Engineering}

Teruel \etal{} \cite{teruel2011csrml} define collaborative systems as systems that allow users to work cooperatively while performing tasks. Whitehead \cite{whitehead2007collaboration} states that stakeholders engage in a range of activities throughout the software development process including the creation of software artifacts such as source code, models, documentation, and test scenarios. Therefore, coordinating numerous stakeholders' efforts is essential to form a viable environment as well as producing more quality software products. However, collaboration is challenging in such environments consisting of stakeholders with different backgrounds and even different cultures in distributed software projects. 

Constantino \etal{} \cite{constantino2020understanding} claim that distributed software projects suffer from several issues in terms of collaboration and also point out that despite these challenges stakeholders in collaborative environments produce better artifacts compared to any single developer's effort. Herbsleb \cite{herbsleb2007global} also states that distributed development environments are negatively affected in terms of coordination and discusses the activities designed to provide coordination among stakeholders. Collaboration in distributed software development environments is provided by various tools to automate and facilitate the entire development process. Lanubile \etal{} \cite{lanubile2010collaboration} state that collaboration tools let stakeholders work together even when stakeholders work apart from each other and provided a list of collaborative development tool categories as (i) version-control systems, (ii) trackers, (iii) build tools, (iv) modelers, (v) knowledge centers, (vi) communication tools and (vii) Web 2.0 applications.

Konaté \etal{} \cite{konate2014collaborative} state that requirements engineering procedures are interdisciplinary processes that necessitate specialized skills from all stakeholders and draw attention to the importance of collaboration in requirements engineering since a single stakeholder cannot satisfy all of the skills required for the job. Azadegan \etal{} \cite{azadegan2013collaborative} state that stakeholders in teams need to work collaboratively and also point out the fact that each individual has a different background and perspective. Even though maintaining a collaborative environment is difficult, collaborative methods in requirements engineering produce better results. With the emergence of crowd-based requirements engineering (CrowdRE), collaboration becomes more crucial since the essence of CrowdRE relies on employing larger numbers of individuals to increase user involvement. Hosseini \etal{} \cite{hosseini2015configuring} state that coordination becomes a problem when requirements engineers have to engage with a big group of users and argue that a major challenge is developing software-based mechanisms to coordinate the crowd with minimal developer participation while being cost-effective.

Literature shows that collaboration is a key concept for software engineering as well as requirements engineering. However, maintaining a collaborative environment is challenging due to the diversity that can be observed among agile teams. We aim to reduce the difficulties of collaboration by providing suggestions to stakeholders depending on their effort as well as their effort as a team so that \sys{} mitigates the risks that might arise due to defective interaction between stakeholders.

\subsection{NLP for Requirements Engineering} 

Applying requirements engineering techniques in a software project is crucial since a great portion of the projects throughout history either have failed or exceeded the budget due to the lack of proper requirements definitions \cite{stallinger2001system}. Davis \etal{} \cite{davis2006effectiveness} prove that software requirements definitions affect the quality of the product significantly. Lamsweerde \cite{van2000requirements} points out the increasing importance of requirements engineering in software engineering since defining requirement properly is a challenging task.

Dalpiaz \etal{} \cite{dalpiaz2018natural} state that natural language (NL) is widely used in requirements engineering (RE) due to the ease of understanding even by inexperienced stakeholders besides the ambiguity of NL. It can be inferred that NLP complements RE so that a detailed analysis might be conducted. The close relationship between the fields of NLP and RE leads to the field of NLP4RE emerging. Zhao \etal{} \cite{zhao2020natural} define NLP4RE as "an area of research and development that seeks to apply NLP techniques requirements artifacts". There are numerous studies in the literature in the area of NLP4RE. Ferrari and Esuli \cite{ferrari2019nlp} propose a method for detecting cross-domain ambiguities in RE by building domain-specific language models to approximate the potential ambiguities. Duan \etal{} \cite{duan2009towards} propose a method for automated requirements prioritization technique based on clustering requirements by their probability of distribution and co-occurrence. Robeer \etal{} \cite{robeer2016automated} propose the Visual Narrator tool that generates conceptual models from user stories by combining several NLP heuristics. Hotomski and Glinz \cite{hotomski2019guidegen} propose the GuideGen approach that generates natural language guidance when the acceptance test needs to be aligned with the requirements definitions. Gemkow \etal{} \cite{gemkow2018automatic} propose a method for glossary term extraction from large-scale requirements documents by using several NLP techniques and statistical filtering. 

Recent studies show interest in applying deep learning and machine learning methods in the field of RE due to the low generalization of traditional methods. Stanik \etal{} \cite{stanik2019classifying} proposes a comparative approach between supervised machine learning (ML) and deep learning (DL) methods by applying these methods to a user feedback classification task and reported that supervised ML performed slightly better than the deep learning method due to the limited number of training samples. By their nature, deep learning methods require a larger number of training samples compared to supervised ML models. However, pre-trained models seem to provide a solution to this problem. Hey \etal{} \cite{hey2020norbert} proposes a fine-tuned version of the BERT deep language model as NoRBERT for requirements classification and reported highly promising results in classification tasks in the non-functional requirements dataset. Araújo \etal{} \cite{de2021re} also propose a RE-specific version to extract requirements from app reviews and reported that their fine-tuned version of the BERT language model outperforms all of the existing methods.

Surveyed literature shows that NLP4RE is an emerging and widely studied area. Furthermore, pre-trained models convey a high potential for several tasks in requirements engineering. We aim to contribute NLP4RE area by applying an NLP pipeline that employs a pre-trained deep language model.

\subsection{CrowdRE and Gamification}

User involvement has favorable consequences on system success and it is shown that employing users as an information source is an effective way to elicit requirements \cite{kujala2003user, kujala2005role}. According to El Emam \etal{} \cite{el1996user}, more user participation reduces the detrimental impact of ambiguity on the quality of requirements engineering tasks. 
Kujala \etal{} \cite{kujala2005role} point out that increasing user involvement in requirements elicitation activities yields well-formed requirements so that the chance of project success is increased. Crowdsourcing is a scalable and inexpensive solution that increases user involvement in RE processes. \cite{snijders2015refine, snijders2014crowd} 

However, providing healthy interaction between users is challenging. To increase user interaction towards systems, gamification is widely applied in many areas such as finance, education, and health \cite{deterding2011game}. Gamification is also applied in the area of requirements engineering. Snijders \etal{} \cite{snijders2014crowd} state that gamification complements crowdsourcing by motivating users to participate more by rewarding the users. Fernandes \etal{} \cite{fernandes2012ithink} propose the gamified environment iThink, to support collaborative requirements elicitation. Snijders \etal{} \cite{snijders2015refine} propose REfine online platform that is enriched with gamification techniques to facilitate requirements elicitation tasks. 

Surveyed literature shows that involving greater numbers of users increases the number and quality of requirements. However, the collaboration between users and user interaction with the system becomes more difficult with the increasing number of users. Our method provides an easy-to-use user interface and several functionalities such as highlighting user stories with detected problems to increase user interaction with the system. In future work, we plan to employ additional gamified objects in \sys{}.

\section{Research Approach}
\label{section:research_approach}

Although user stories are widely adopted in the industry \cite{lucassen2016use,kassab2015changing,cohn2004user} and studied in academia \cite{dalpiaz2018agile, lucassen2015forging, dalpiaz2019detecting, barbosa2016use}, existing work focus on the user stories as a set by either checking the quality \cite{wake2003invest, lucassen2015forging, lucassen2016improving} or extracting models from the final set \cite{trkman2019impact, gunecs2020automated, wautelet2016building, jaqueira2013using, wautelet2018modelers}. We identify a gap in the literature regarding collaboratively building a user story set rather than analyzing a complete set of user stories. This discovery leads us to our main research question: 

\paragraph{\textbf{MRQ.} How can we support stakeholders in a collaborative environment while constructing user story sets?}

\begin{figure}[htbp]
\centering
  \includegraphics[width=\linewidth]{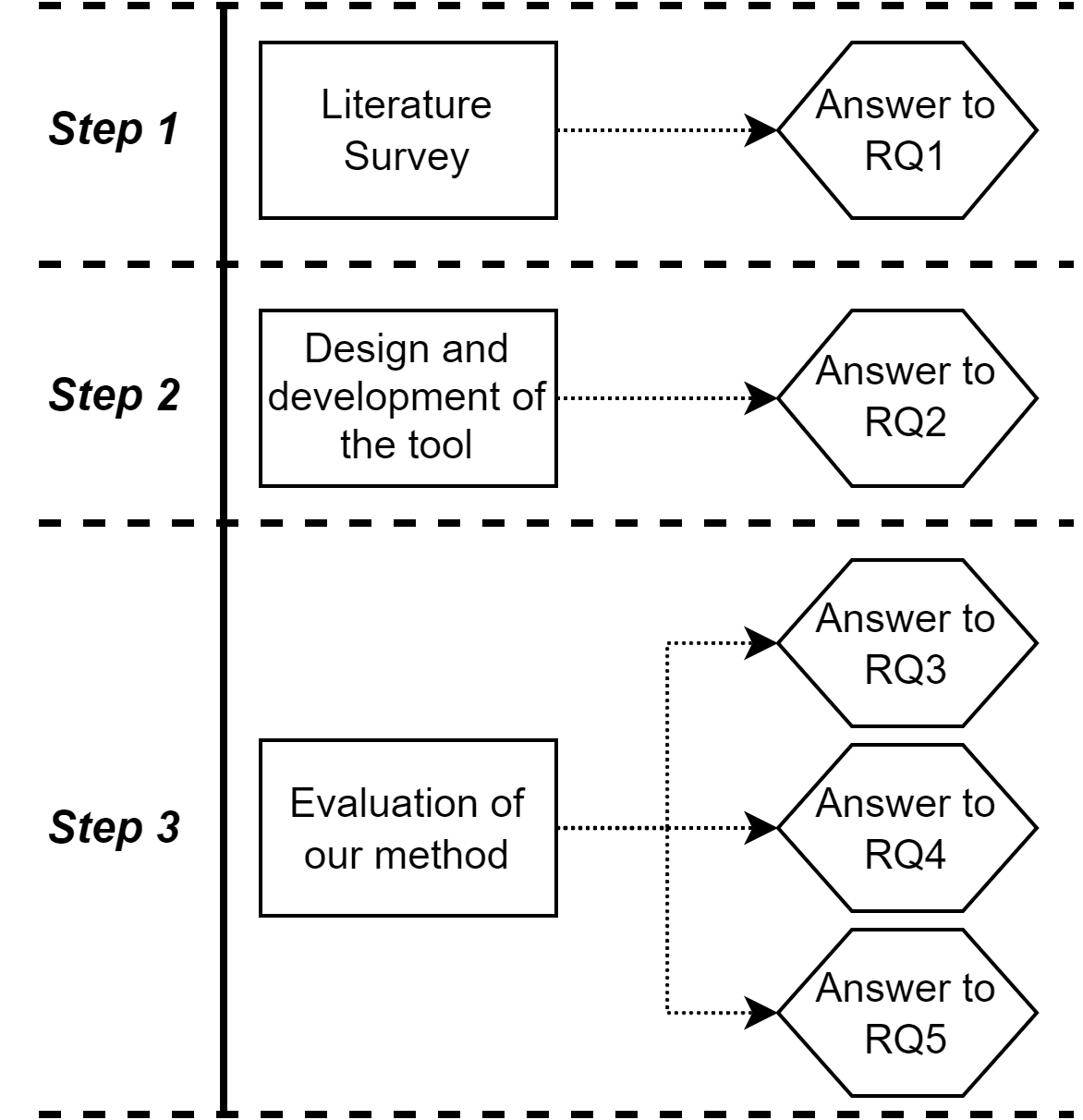}
  \caption{Steps of our research approach}
  \label{fig:research_approach}
\end{figure}

To address this question, we design our study following Wieringa's three-step cycle of the design science research approach \cite{wieringa2014design} as summarized in Figure~\ref{fig:research_approach}. 

To identify the issues with the status quo, we conduct \textit{problem investigation}. This creates our first research question.
\paragraph{\textbf{RQ1.}What are the issues of the current agile requirements engineering practices?}
We conduct a thorough literature survey to reveal these limitations in different aspects. Surveyed literature clearly states,

\begin{table}[ht!]
\centering
\caption{Issues of the current agile requirements engineering practices}
\begin{tabular}{ll}
\toprule
Issue ID & Description \\
\midrule
Issue 1 & Difficulties arise from natural language \\ 
Issue 2 & Effort required for fixing defects \\ 
Issue 3 & Need for better tools for collaboration \\ 
\bottomrule
\end{tabular}%
\end{table}

\begin{enumerate} 
  \item Even though the widespread usage of NL requirements representations, practitioners often tend to construct error-prone solutions even with simple structured notations such as user stories due to the ambiguity that arises from the nature of natural language. \cite{kassab2014state, lucassen2016use, kassab2015changing, cohn2004user, lucassen2015forging} To control the quality of user story sets, several methods are proposed but they require practitioners to be trained. \cite{doe2011recommended, heck2014quality, wake2003invest, lucassen2015forging, lucassen2016improving}
  \item Several methods that employ NLP techniques are proposed to detect defects in requirements documents. \cite{dalpiaz2019detecting, yang2011analysing, barbosa2016use, yang2010automatic} To reduce the time spent fixing defects in requirements documents, a method that helps practitioners to improve user stories as they are written, is needed.
  \item Software development processes and requirements engineering practices require stakeholders to work cooperatively on different tasks to produce better output \cite{teruel2011csrml, whitehead2007collaboration, constantino2020understanding, herbsleb2007global, lanubile2010collaboration}. Requirements engineering practices are interdisciplinary processes and the emergence of CrowdRE requires a large number of practitioners to work collaboratively \cite{konate2014collaborative, azadegan2013collaborative, hosseini2015configuring}. However, the increasing sizes of agile teams along with members in separate locations make preserving viable collaborative environments harder. Software-based smart tools can be developed to provide a potential solution to this issue.
  
\end{enumerate}
With this motivation at hand, we aim to improve the status quo concerning 
the second step of Wieringa’s design cycle, \textit{design a treatment}. This creates our three subsequent research questions.
\paragraph{\textbf{RQ2.} What is an effective artifact that can support stakeholders when constructing user stories?}
We propose \sys{}, a collaborative requirements editor that supports stakeholders by generating on-demand suggestions. Our method is detailed in the rest of the paper.
We decided to create a tool that has a simple UI that allows users to interact with the system easily. On top of this, we instrument \sys{} with a state-of-the-art NLP pipeline that employs a pre-trained deep language model.

Finally, we perform \textit{treatment validation} for the design artifact as suggested by Wieringa’s design cycle. The following three questions address our \textit{treatment validation} phase. 

\paragraph{\textbf{RQ3.} How effective is our system in terms of ensuring the completeness of the user story set?} 

The efficient use of resources in large-scale agile teams greatly depends on the level of collaboration among the stakeholders yet maintaining collaboration is a challenging task in distributed environments \cite{stallinger2001system, constantino2020understanding, herbsleb2007global}. Collaboration needs to be supported with different instruments. When it comes to requirements engineering, practitioners with diverse skills put effort to satisfy the goals of the project. \cite{konate2014collaborative, azadegan2013collaborative} An increased level of collaboration reflects positively on the quality of the requirements artifacts \cite{kujala2003user, kujala2005role}. However, the collaborative environment needs to be properly maintained otherwise several defects such as incompleteness might occur in a user story set \cite{lucassen2015forging}.

To increase the completeness of the user story sets, we benefit from the collaborative input of the participants to generate our suggestions. We analyzed the number of user stories and standard deviation of user stories along with the number of isolated concepts that are present among user stories. An increased number of user stories and a reduced number of isolated concepts have a positive impact on the completeness of the user story set.  Also, we treat the graphs that are generated via our graph generation module as trees and we applied \textbf{breadth first search (BFS)} traversal. The increasing number of nodes in BFS traversal indicates that users provide more detailed properties for each concept that is supported with different sentences.

\paragraph{\textbf{RQ4.} To what extent do the participants implement the suggestions produced by \sys{}?}

\sys{} aims to increase the completeness of the user story sets by providing suggestions to stakeholders. \sys{} implements different heuristics to generate suggestions for users. Thusly, we assess the success of the heuristics as well as \sys{} itself, by measuring the level of incorporation of the generated suggestions. We generate different suggestions for the participants throughout the experiment sessions. We expect users to apply some of these suggestions that contribute to their way of thinking. The adoption ratio of the suggestions is directly proportional to the effectiveness of \sys{}. This RQ aims to measure the portion of the suggestions that are accepted as useful and applied by the participants. This enables us to measure the effectiveness of our suggestion strategies.

\paragraph{\textbf{RQ5.} What is the perceived usability level?}

We directly address an industry problem hence \sys{} must be useful for requirements engineering professionals. Therefore this RQ aims to understand whether users consider \sys{} beneficial. The participants were asked several questions on a 5-point Likert scale in the post-experiment survey to evaluate the effectiveness of \sys{}, the user interface, and the level of usefulness of provided suggestions in terms of quality and completeness.

\section{\sys{}: Smart Collaborative User Story Tool}
\label{section:method}

\begin{figure}[htbp]
\centering
  \includegraphics[width=\linewidth]{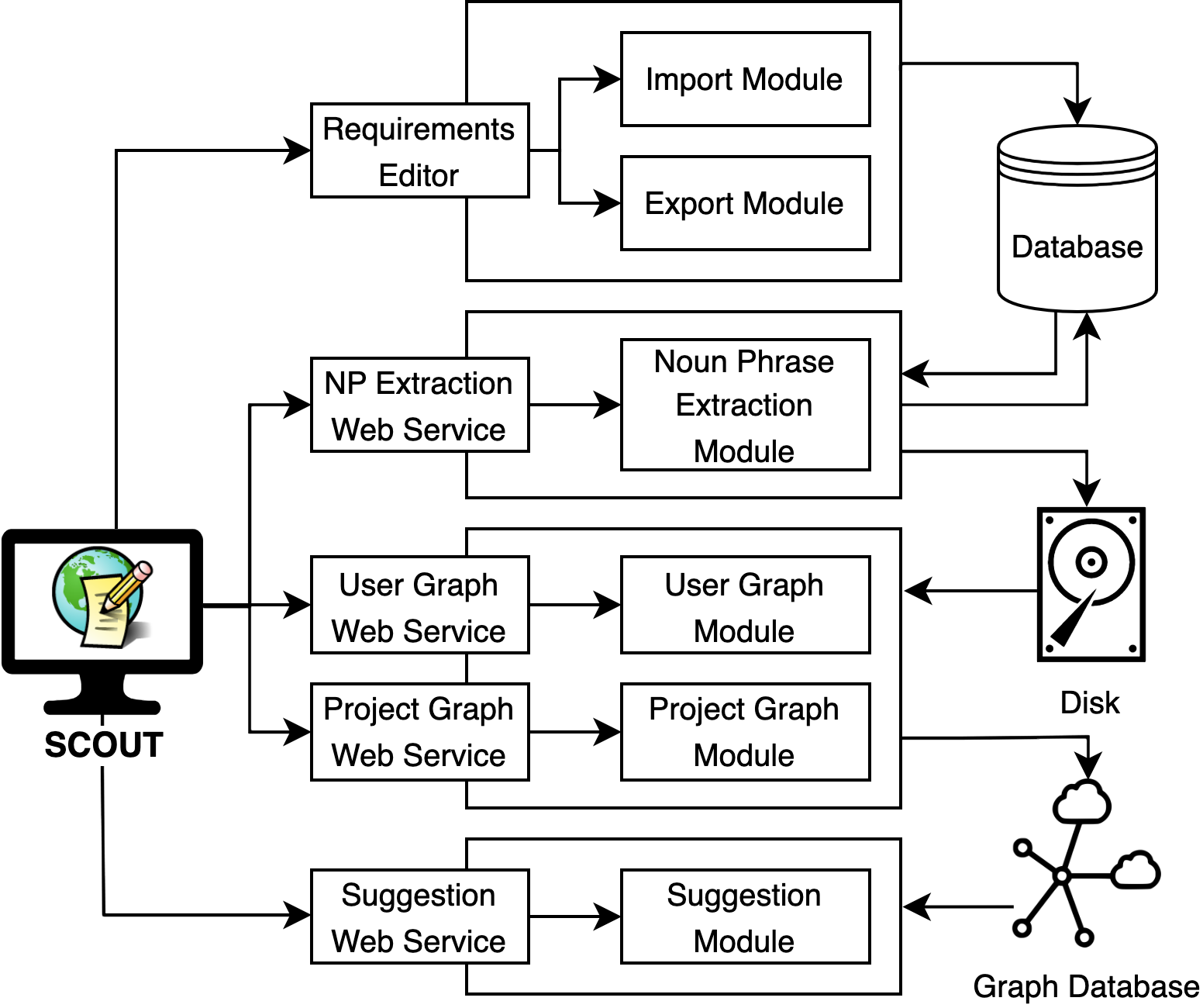}
  \caption{Architecture of the \sys{}}
  \label{fig:architecture}
\end{figure}

Our goal in this work is to create a web-based collaborative agile requirements tool that supports stakeholders as they create user stories in a shared environment. We develop \sys{} to achieve this goal. Stakeholders can create and store user stories with \sys{} so they can use them to gather suggestions and make the user story set more complete. Stakeholders may be assigned to teams and projects so they can carry out various tasks for various projects. Both the individual efforts of stakeholders and the whole set of user stories for a project that is created by all stakeholders are benefited by \sys{}. To create an environment that is viable for stakeholders, we implement various modules and submodules. Web services are used to provide a communication channel between the editor and the modules. Fig. \ref{fig:architecture} demonstrates the architecture of \sys{}.

\paragraph{Collaborative Agile Requirements Editor} We implement a collaborative agile requirements editor for stakeholders. We instrument our interface with two different sub-modules as import and export modules. These modules serve to increase the usability of the system. Stakeholders can easily import user story sets in hand and export user story sets into different file formats. The user interface also allows stakeholders to gather suggestions with one click, and chat with other stakeholders that work on the same project. 

\paragraph{Noun Phrase Extraction Module}
We implement a noun phrase extraction module as an initial step to extract the most informative parts of user stories. Our noun phrase extraction module is responsible for processing raw natural language text in user stories into clean noun phrases and transferring the extracted information to the database.

\paragraph{Graph Generation Module}

We aim to transform the output of the noun phrase extraction module into the concepts present in the user story set using the relatedness between the extracted noun phrases. To achieve this goal, we implement an NLP pipeline that depends on pre-trained BERT sentence embeddings. After we extract concepts from noun phrases, we store these concepts and their relations in a graph database. We use this information in the graph database for visualization and further steps of our implementation. We implement two different web services to populate the graph database for each stakeholder and project.

\paragraph{Suggestion Module}

We aim to generate suggestions to increase the completeness of the user story set. To achieve this goal, we benefit from the relations between concepts that are stored as knowledge graphs. We implement different heuristics to generate suggestions for stakeholders using individual stakeholders' input along with all stakeholders' input for a project.

%

\begin{figure*}[htbp]
    \centering
    \includegraphics[width=\textwidth]{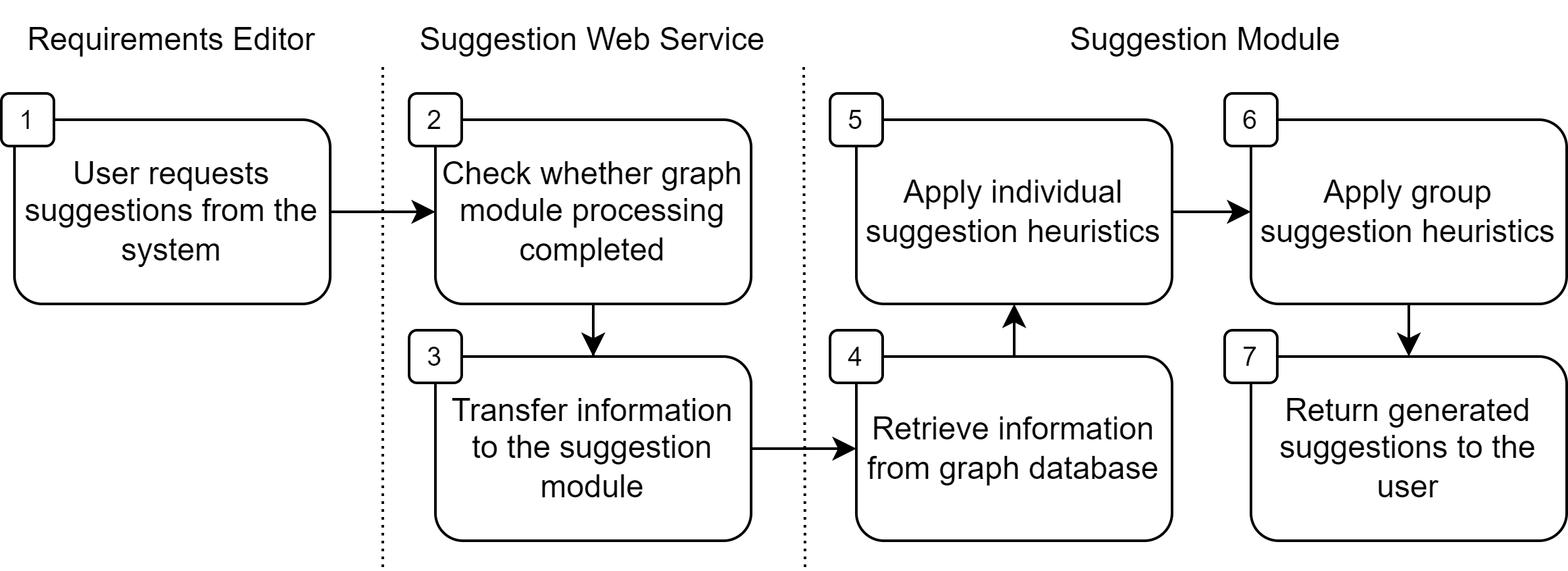}
    \caption{Pipeline of inner workings of suggestion module}
  \label{fig:architecture_sug_mod}
\end{figure*}

The internal workings of \sys{}'s noun phrase extraction, graph, and suggestion modules are shown in  Fig. \ref{fig:architecture_np_extr}, Fig. \ref{fig:architecture_graph_mod}, and Fig. \ref{fig:architecture_sug_mod}, respectively.


\section{Collaborative Agile Requirements Editor}
\label{section:editor}
\begin{figure*}[htbp]
    \centering
    \includegraphics[width=\textwidth]{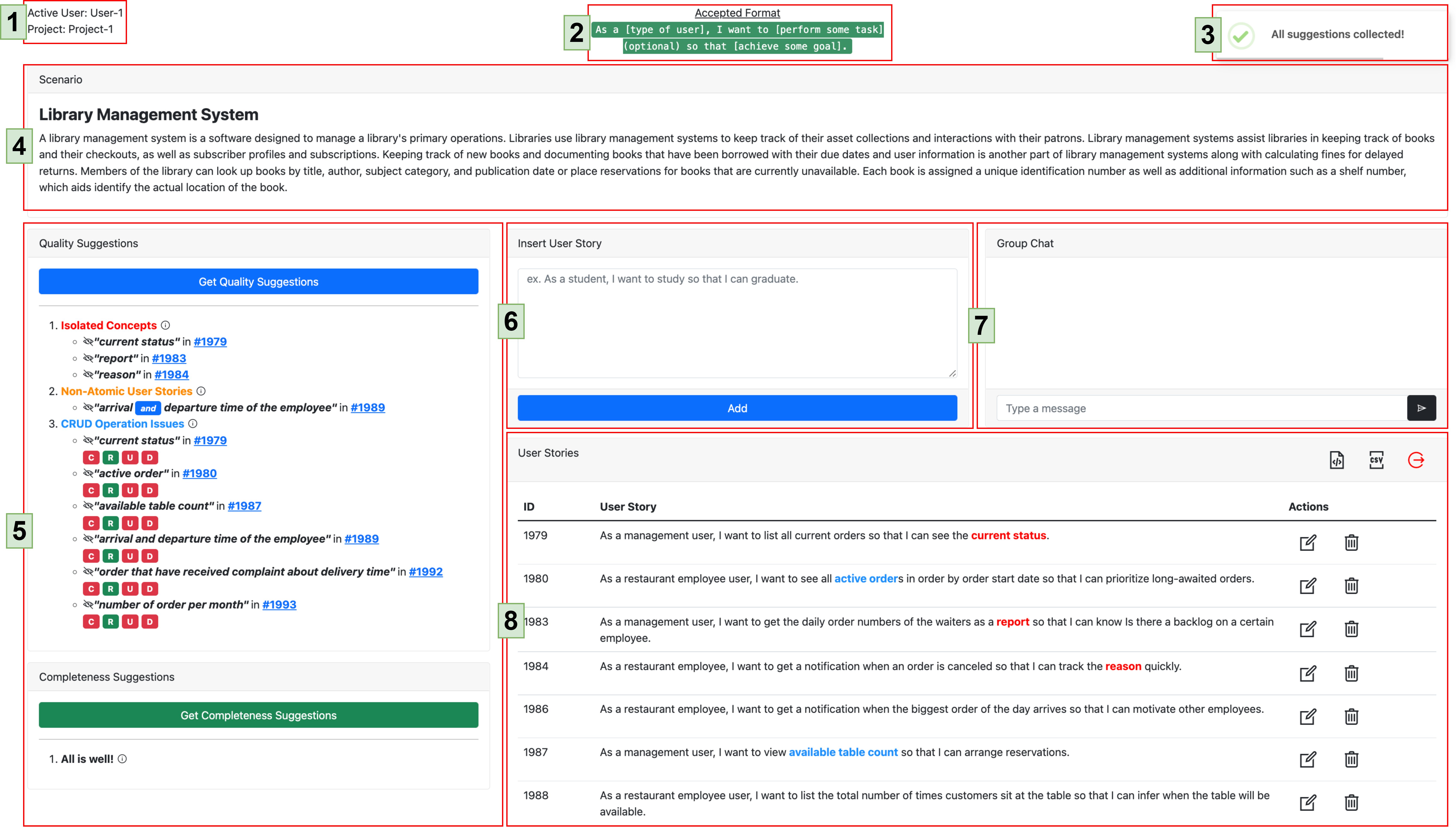}
    \caption{User interface of the collaborative requirements editor of \sys{}{}}
    \label{fig:ui}
\end{figure*}

We implement an interactive interface that increases user interaction with the system. Fig. \ref{fig:ui} presents a screenshot of the interface of our system. We explain the different user interface components (UC) as follows,

\textit{UC \#1}. This component shows the information about the user that is logged in and which project the user is currently working on.

\textit{UC \#2}. This component shows the accepted user story format. To eliminate the errors that might occur because of the sentences that do not follow the user story format. We add this component on top of the main page which helps users quickly control whether the sentence complies with the format. We also check the format with a regular expression if the user does not enter a valid user story an error message is shown to the user and the entered sentence will be available in UC \#5 for editing.

\textit{UC \#3}. This component shows different messages for different actions that can be performed by the user. It informs users whether the action they perform results in success or failure. 

\textit{UC \#4}. This component holds the scenario text associated with the project. We put the scenario text in the middle of the screen to enable users to check that text without any disturbance.

\textit{UC \#5}. This component is used to request and display suggestions. Users can request quality and completeness suggestions from \sys{}. These suggestions are listed in different areas in this component. \sys{} provides three types of quality suggestions as isolated, atomic and CRUD suggestions and seven types of completeness suggestions as close to completeness, pop-zero, pop-one, pop-two, pop-three, feeling lucky and all is well to the users. Detailed information about generated suggestion types is provided in the Sec. \ref{section:suggestion_module}. Suggestions are displayed in this panel and divided into separate categories. Users can report suggestions that are not useful for their purpose by clicking on the crossed-eye icons. Users also can easily navigate the sentences that are related to a quality suggestion by clicking on the hyperlinks at the end of each quality suggestion.

\textit{UC \#6}. This component is used to add new user stories to the system. If the added user story does not comply with the user story format, this component will be automatically filled with the sentence again to allow the user to quickly fix that sentence.

\textit{UC \#7}. This component is used to chat with other members of the same team. Sent messages are displayed in real-time for all users of the same team.   
\textit{UC \#8}. This component is mainly used to display user stories that are constructed by the logged-in user. Users can edit and delete user stories. Users can also export these user stories as JSON and CSV formats using the buttons in the upper right corner. As can be seen in Fig. \ref{fig:ui}, some of the user stories have colorful parts in them. These parts are generated according to the retrieved suggestions to pinpoint the parts in sentences that might be fixed.

\section{Noun Phrase Extraction Module}
\label{section:np_extr_module}
\begin{figure*}[htbp]
    \centering
    \includegraphics[width=\textwidth]{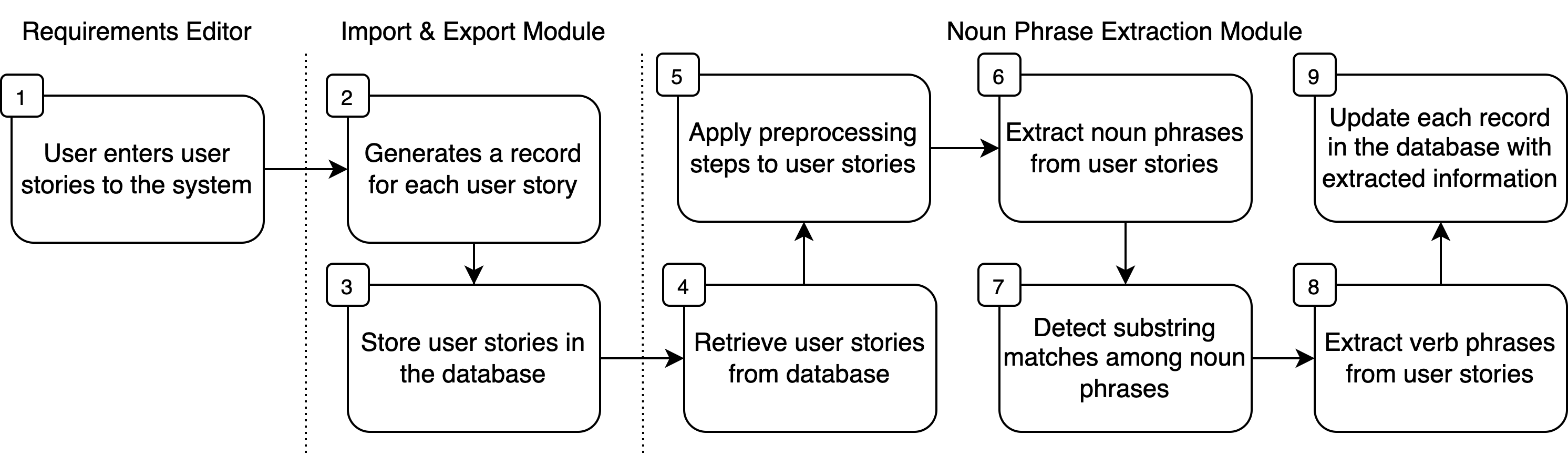}
    \caption{Pipeline of inner workings of noun phrase extraction module}
    \label{fig:architecture_np_extr}
\end{figure*}

The noun phrase extraction module analyzes the natural language text in user stories. Since stakeholders create user stories using unconstrained NL, a list of noun phrases is extracted from the raw NL text via our NLP pipeline. Our noun phrase extraction pipeline is summarized in Fig. \ref{fig:architecture_np_extr} and the steps of our noun phrase extraction module are detailed in the rest of this section.

\textit{Steps 1 - 4: Preliminary steps of Noun Phrase Extraction Module}. Our pipeline starts with user interaction. (Step 1) Users can add single or multiple user stories to the system from the user interface. (Step 2) With the help of the import module, we parse the user input and create records that involve user, project, and time information for each record. After that, we store these records in a database. We also provide an export module for users to export their input as JSON or CSV formats. (Step 3) Our noun phrase extraction pipeline starts by retrieving these user stories by querying the database. (Step 4)

\textit{Step 5: Pre-processing}. We use several pre-processing techniques to enhance the extraction of noun phrases from user stories.  Our noun phrase extraction module includes several steps, such as (i) case-folding to normalize the user stories. (ii) punctuation removal to eliminate unnecessary information, (iii) lemmatization to reduce inflectional forms of words. Due to stemming's reliance on vocabulary and morphological aspects of words, we prefer lemmatization. Stop-word removal is widely used in NLP tasks \cite{zhao2020natural}. Nevertheless, following several test runs, we decide against removing stop words from the user stories. Even though stop-word removal generally leads to better results for NLP tasks, in our case doing so produces fairly poor results because the absence of conjunctions and prepositions results in parts of sentences' merging unintentionally. This produces meaningless noun phrases that affect our other steps of implementation. We put these steps into implementation by utilizing NLTK and native Python libraries to discover the concepts that exist in user stories and to create a knowledge graph. 

\textit{Step 6: Noun Phrase Extraction}. Our initial method for capturing the information in the user stories is to extract nouns as concepts from the pre-processed user stories after they have been generated. However, because of the inadequate information provided by nouns alone, working exclusively with nouns does not produce satisfactory results. Therefore, we choose to use the spaCy\footnote{https://spacy.io} to extract noun phrases from user stories. For each user story, we extract dependency trees and noun chunks. We combine each extracted noun chunk with its syntactic descendants\footnote{https://spacy.io/api/token} to broaden the scope of our noun phrase extraction module. To accomplish this, we concatenate all of the syntactic descendants of each extracted noun chunk from the leftmost edge to the rightmost edge of the dependency tree.

\textit{Step 7: Substring Matches}. We discovered that there are many noun phrases with parts in common after extracting noun phrases from user stories. They are what we refer to as substring matches that are observable between noun phrases. These noun phrases with parts in common can be grouped using the subset of noun phrases that contain a specific noun phrase as a substring. This process can be simply described as noun phrase clustering. For instance, the noun phrases "item label" and "item label size" can be grouped under the "item label" key since "item label size" refers to a property of "item label". By using this step, all of these noun phrases are grouped into a set, with the shortest noun phrase serving as their representation. The number of noun phrases in the initial set is decreased by applying this step. The initial size of the noun phrase set being reduced also reduces the time and space complexity of subsequent steps in our pipeline.

\textit{Step 8: CRUD Operation Extraction}. We discover that verb phrases in user stories also contain important information after extracting noun phrases. We develop a sub-module in our pipeline for extracting noun phrases that extract this data from verb phrases. Our preliminary tests, however, show that not all verb phrases provide information that is pertinent to our goal. We develop the concept of verb phrase extraction for CRUD (create-read-update-delete) operations that can be used on an artifact after conducting a thorough investigation. The completeness of the user story set could be jeopardized by the absence of coherent CRUD operations. For instance, unless it has been created beforehand, no project artifact can be deleted. As shown in the example, this technique satisfies our initial goal of making the user story set more complete. However, this operation cannot be performed with just four verbs, so we must develop a glossary of terms that can be used in place of the four primary operations. Our glossary was created using a thesaurus found online\footnote{https://www.thesaurus.com/}. After that, we extract verbs from user stories using the pattern-matching technique of textacy\footnote{https://github.com/chartbeat-labs/textacy}. Then, to extract complete verb phrases from user stories, we navigate the syntactic descendant tree. After retrieving verb phrases, we divide them into four main categories referred to as CRUD.

\textit{Step 9: Updating Records}. After applying all of the steps of our noun phrase extraction pipeline. Our noun phrase extraction module updates each record with the extracted information from the user stories.



\section{User \& Project Graph Modules}
\label{section:graph_module}

\begin{figure*}[htbp]
    \centering
    \includegraphics[width=\textwidth]{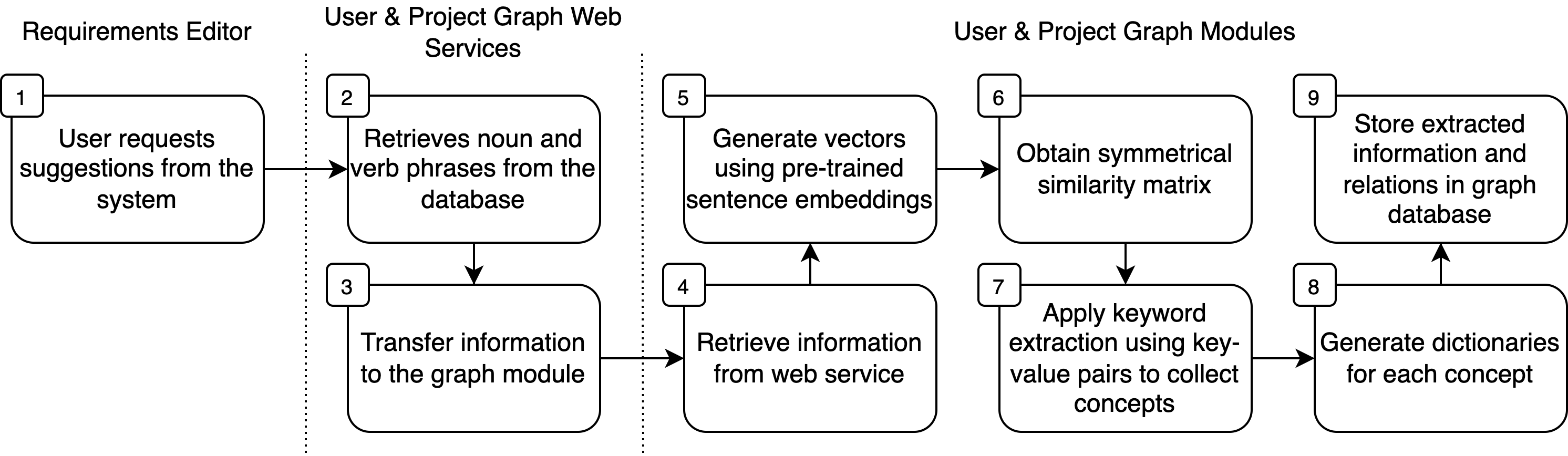}
    \caption{Pipeline of inner workings of graph modules}
  \label{fig:architecture_graph_mod}
\end{figure*}

We implement a graph module that uses an NLP pipeline powered with a deep language model to extract concepts from the output of the noun phrase extraction module and store this data in a graph database. To process the work of individual stakeholders as well as the entire project user story set, which was built by numerous stakeholders, we develop two distinct web services. Our graph module uses the extracted concepts and their relations to populate two different graph databases. Our graph module is summarized in Fig. \ref{fig:architecture_graph_mod} and the steps of our graph module are detailed in the rest of this section.

\textit{Steps 1 - 4: Preliminary steps of Graph Modules}. Our pipeline starts with user interaction. Users can request suggestions from the interface of \sys{}. (Step 1) This request is processed by our user \& project graph web services which act as our middleware layers. These services retrieve concepts to the graph module. (Step 2)  The extracted information is transferred to the user \& project graph modules. (Step 3) Our graph module starts by retrieving the information from the web service. (Step 4)

\textit{Steps 5 \& 6: Sentence Embeddings}. In the noun phrase extraction module, we are able to group a few of the user stories although independent user stories still exist. Although requirements engineering experts created the user stories, experts with diverse backgrounds may use different words to describe the same idea. For instance, due to their close resemblance, the noun phrases "shipping offers" and "shipment options" will be matched as similar noun phrases. We take advantage of BERT, a state-of-the-art deep language model \cite{devlin2018bert}, to reveal this relation and broaden the scope of concept extraction. As a result, knowledge graph extraction from user stories is made possible by comparing the similarity of various noun phrases to one another and by identifying various noun phrases that refer to the same meaning using pre-trained sentence embeddings.

First, Hugging Face's\footnote{https://huggingface.co} sentence transformers, which are intended to be used with a trained model, are used to encode noun phrases. For sentence embedding, many pre-trained sentence embedding models are available. We select \textit{paraphrase-mpnet-base-v2} because of its great performance on the STSbenchmark test. For each noun phrase, this sentence encoding process produces vectors with 768 dimensions. Pairwise cosine similarities between each noun phrase are calculated to extract similar noun phrases, and the symmetrical similarity matrix is then obtained.

\begin{algorithm}[htp]
\SetAlgoLined
\SetKwInOut{input}{Input}
\SetKwInOut{output}{Output}
\input{a dictionary L of noun phrases and a matrix S of pairwise similarities}
\output{a dictionary D of similar noun phrase pairs}
D = \{\}\;
threshold = 0.4\;
\ForEach{term T in keys of L}{
similarity\_score = -1\;
P = ''\;
terms = list of terms in the similarity matrix S\\
\ForEach{term N in terms}{
\uIf{T is not equal to N}{
score = similarity score between T and N\\
  \uIf{score $>$ similarity\_score and score $>$ threshold}{
   similarity\_score = score\;
   P = N\;
  }
}
}
P = keywordExtraction(P, T)\;
\uIf{P is not empty}{
append T to D[P]\\
}
\Else{append P to D[T]\\}
}
append dictionary L into dictionary D\\
\ForEach{key K in keys of D}{
\uIf{D[K] is empty}{
delete D[K]
}
}
\textbf{return} D\;
 \caption{Sentence Embeddings}
\end{algorithm}

The goal is to collect the noun phrase pairs with the highest similarity after obtaining the similarity matrix. To achieve this, we create an algorithm that iteratively creates pairs of noun phrases that are the most similar to one another. The number of candidate similar noun phrases is limited with a pre-defined similarity threshold to reduce the number of iterations necessary. The predefined similarity threshold is set to 0.4 after several test runs. (Line 2) The similarity score between a noun phrase and each candidate noun phrase is taken from the similarity matrix during each iteration and compared with a noun phrase that was previously chosen and the target noun phrase. (Lines 7 - 9) Only in cases where the new noun phrase's similarity score exceeds both the threshold and the previous similarity score will the target noun phrase and similarity score be updated. (Lines 10 - 12)

There may be a smaller common unit of representation for the noun phrases in the key-value pairs, even though the majority of similar key-value pairs are extracted from the noun phrases. With a pre-trained sentence embedding model, KeyBERT \cite{grootendorst2020keybert} from Hugging Face is used to extract keywords from the key-value pairs to provide a higher level of abstraction. We selected \textit{paraphrase-mpnet-base-v2} because of its great results on both our sentence embedding step of the graph module and the STSbenchmark test. (Line 14)

\begin{algorithm}[htp]
\SetAlgoLined
\SetKwInOut{input}{Input}
\SetKwInOut{output}{Output}
\input{parent term P and term T}
\output{parent term}
parent = P\;
term\_string = T + parent\;
keywords = extracted keyword and probability pairs from term\_string using KEYBERT\\
\uIf{length of keywords $>$ 1}{
set the most probable keyword as parent\\
}
\textbf{return} parent\;
 \caption{Keyword Extraction}
\end{algorithm}

\textit{Step 7: Keyword Extraction}. To extract the most representative n-grams from the texts, keyword extraction techniques are used. In this step, lemmatized versions of each pair's key and value are gathered and joined together to create a single string (Line 2). KeyBERT is adjusted to extract unigrams from that concatenated string  (Line 3). Both the key and the value are grouped under the new representation if there is a smaller common unit of representation for a key-value pair. (Lines 4 - 5)

\textit{Step 8: Generating Dictionaries}. Related and parent terms are kept in a dictionary after using keyword extraction. The parent terms will be the dictionary's keys, and the terms that are related to the parent terms will be their values. Each term is added to the dictionary as a value of the parent term. (Lines 15 - 16) When a term lacks a related term with a similarity score greater than the threshold, we append it as a parent term, and its value is set as a list with an empty string. They will be handled as separate terms with no relation to other terms. (Lines 17 - 18). The output of the noun phrase extraction module is added to the dictionary that was created during the iteration. (Line 21) To get rid of terms that the noun phrase extraction module did not process, we delete key-value pairs with empty values. (Lines 22 - 24)

\textit{Step 9: Graph Database}. A graph database is used to store all concepts and the relationships between them that are extracted from the user stories. Since identifying the relationships between concepts is our main goal, we decided to store the extracted concepts in a graph database because doing so enables us to gain insightful information. To store our data, we choose the open-source Neo4j graph database \footnote{https://neo4j.com}, and to access the database from our graph module, we employ the Neo4j Bolt Driver \footnote{https://github.com/neo4j/neo4j-python-driver}. We build two distinct databases for user and project graphs so that we can examine the user story set from various angles. For both databases, we define the same node structure that has the following attributes: (i) key as the concept; (ii) user\_id as the user's unique identifier; (iii) project\_id as the project's unique identifier; (iv) user\_story as the list of user stories related to the concept; (v) is\_active as the flag indicates whether that node is generated by the latest commit; (vi) expiry\_date as the timestamp denotes the time that record becomes inactive. is\_active and expiry\_date properties are aligned with each other. Consider a scenario in that a user requests suggestions for the first time. In that case nodes will be created with \textit{is\_active $=$ 1} and \textit{expiry\_date $=$ 9999-12-31} properties. If the user request suggestions again in that state, before adding new nodes existing nodes with \textit{is\_active $=$ 1} property are converted to \textit{is\_active $=$ 0} and their \textit{expiry\_date} is set to the timestamp of that time. Implementing this logic enables us to keep track of the changes in graphs over time. To identify related concepts as edges between nodes, we define a \textit{RELATED\_TO} relationship that denotes the edges between nodes. 

\begin{figure}[htp]
    \centering
    \includegraphics[width=\linewidth]{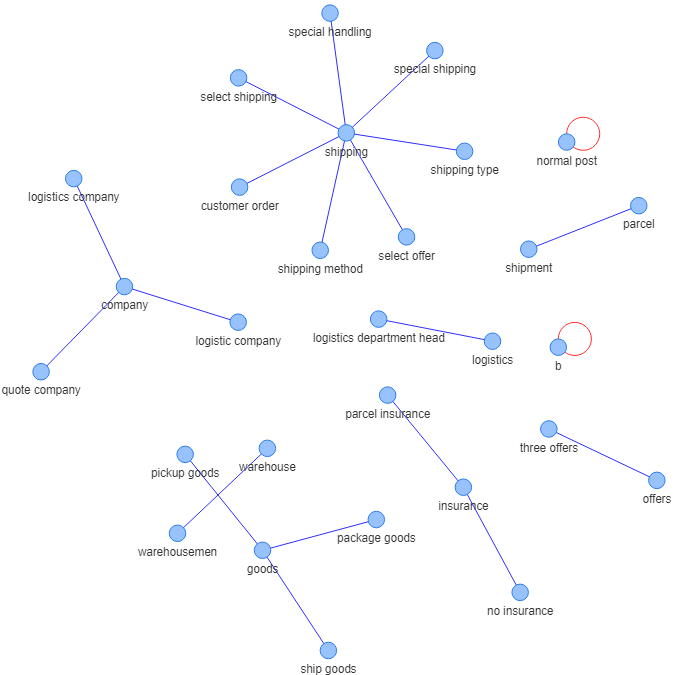}
    \caption{Sample graph representation of extracted terms}
    \label{fig:graph}
\end{figure}

A sample representation of the extracted concepts is shown in Fig \ref{fig:graph}. Red vertices (self-links) indicate mismatched concepts, while blue vertices (matching concepts) indicate this.

\section{Suggestion Module}
\label{section:suggestion_module}
We implement a suggestion module to generate suggestions by applying several heuristics to the information that is stored in the graph databases. We provide nine different types of suggestions divided into categories as quality and completeness suggestions to the stakeholders by querying and traversing the graph databases. Our suggestion module is summarized in Fig. \ref{fig:architecture_graph_mod} and the steps of our suggestion module are detailed in the rest of this section.

\textit{Steps 1 - 4: Preliminary steps of Suggestion Module}. Our pipeline starts with user interaction. (Step 1) After the user requests suggestions from the system. The suggestion web service waits for the graph module to be completed since the suggestion module depends on the output of the graph module. (Step 2) After the graph module is completed, the suggestion web service transfers the information to the suggestion module. (Step 3) Consequently, the suggestion module queries the graph database to gather the required information. (Step 4)

\emph{Step 5: Quality Suggestions}

We aim to assist stakeholders to construct more quality user stories with our quality heuristics. We focus on the user story set of each stakeholder separately when generating quality suggestions. 

\textit{Heuristic 1 - Isolated Concepts}.
We extract the list of concepts from user stories via noun phrase extraction and graph modules. We observe that some of the concepts do not have a relationship with any other concept. We refer to them as self-links and denote them with red edges in our graph representation. In other words, we collect the nodes that only have a relationship with themselves. Unlike other terms, these concepts have the potential for being unrelated when the scope of all user stories that are created by a single stakeholder is taken into account. 

\begin{figure}[htbp]
    \centering
    \includegraphics[width=0.8\linewidth]{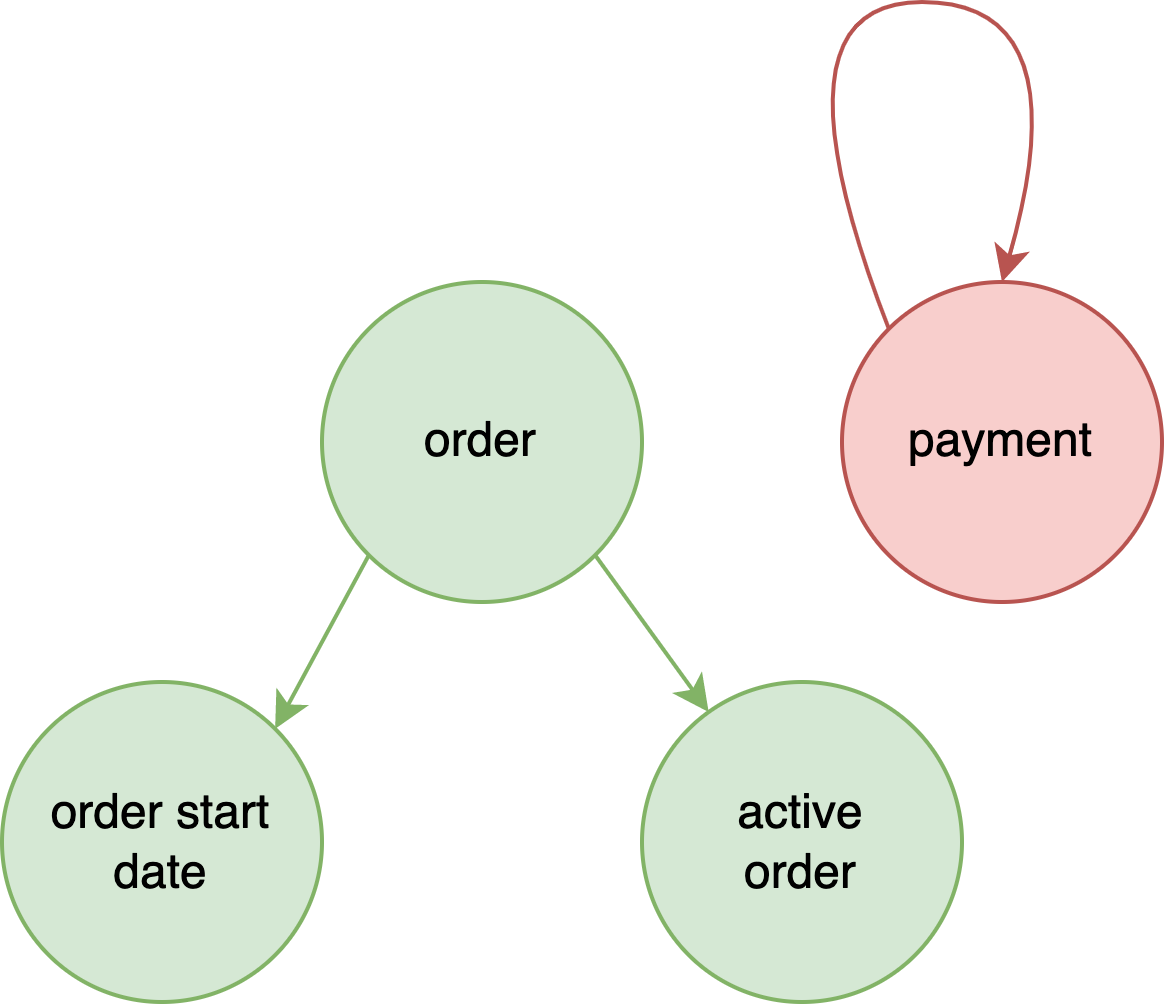}
    \caption{Representation of \textit{isolated concepts} suggestion.}
  \label{fig:suggestions-isolated}
\end{figure}

Our initial aim with this heuristic is to get rid of unrelated terms so that stakeholders will be able to construct unambiguous user stories. To achieve this goal, we label these terms as isolated concepts and suggest stakeholders add more user stories about them or remove user stories about these concepts. Fig. \ref{fig:suggestions-isolated} represents the isolated concepts heuristic.

\begin{figure}[htbp]
    \centering
    \includegraphics[width=\linewidth]{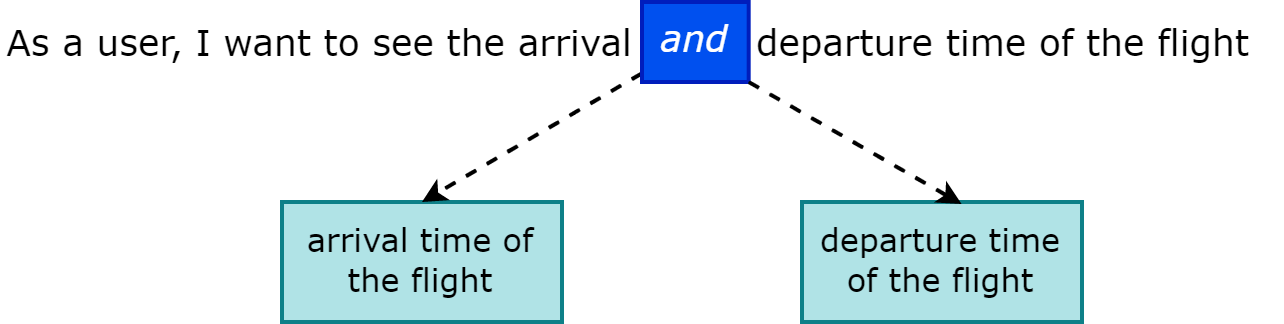}
    \caption{Representation of \textit{non-atomic} suggestion.}
    \label{fig:suggestions-atomic}
\end{figure}

\textit{Heuristic 2 - Non-Atomic User Stories}.
User stories are widely used for describing requirements simply and understandably. Thus, it is commonly accepted that user stories are moderately smaller compared to raw requirements texts. However, stakeholders sometimes tend to use more nouns, adjectives, prepositions, and conjunctions than necessary in user stories. As stated by Lucassen \etal{} \cite{lucassen2015forging}, a user story should specify exactly one feature. Thus, we develop a heuristic that captures noun phrases that contain conjunctions so which eliminates the possibility of breaking the atomicity. By doing so, we manage to avoid misconceptions among stakeholders. We implement this heuristic by capturing the noun phrases that contain the conjunctions and \& or. Fig. \ref{fig:suggestions-atomic} represents the non-atomic user stories heuristic.

\begin{figure}[htbp]
    \centering
    \includegraphics[width=0.8\linewidth]{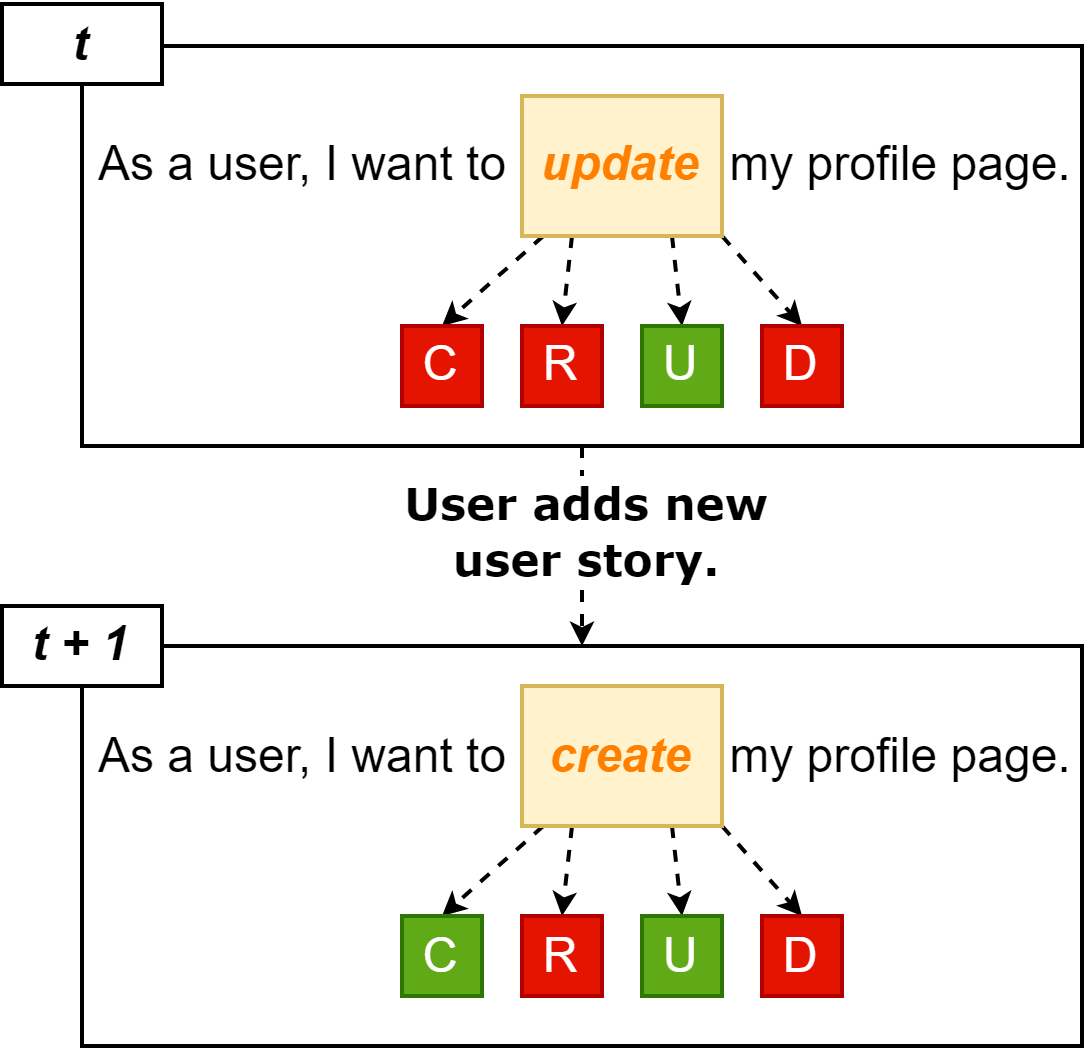}
    \caption{Representation of CRUD suggestion.}
    \label{fig:suggestions-crud}
\end{figure}

\textit{Heuristic 3 - CRUD Operation Issues}.
By their nature, user stories define the actions that can be performed by actors of the software product. In some cases, user stories complement each other when the whole functionalities are considered. It can be inferred that the co-existence of some user stories that define the same part of the software system is a must. For instance, when a stakeholder creates the user story "As a user, I want to view my profile so that I can keep track of my progress.", it can be said that this profile must be created beforehand so that a user can view that profile. Therefore, the action of viewing a profile is dependent on the creation of that profile. We define four main operations as CRUD (create-read-update-delete) and construct a glossary that contains synonym verbs for each operation. We extract verb phrases from user stories to look for verbs that influence the same term. If that term lacks complete CRUD operations, we generate a suggestion for stakeholders to define missing operations for that term.  Fig. \ref{fig:suggestions-crud} represents the CRUD operation issues heuristic.

\emph{Step 6: Completeness Suggestions}

We aim to increase the completeness of the user story set by assisting stakeholders with suggestions. To achieve this goal, we collect all user stories that are constructed for a project by different stakeholders and compare the whole user story set with each stakeholder's user story set to ensure that all stakeholders complement each other to achieve a better user story set with minimal effort.


We benefit from nodes and edges in the graph database that are previously populated via our noun phrase extraction and graph modules. We implement different heuristics that benefit from graph traversal. We extract the most common (top-N) concepts from the project graph as a starting point. We treat them as our main concepts since a vast majority of stakeholders opt for these concepts while constructing user stories. To limit the number of suggestions that will be generated, we limit the number of main concepts to 5. With the main concepts for a project at hand, we extract the main concepts for the stakeholder that requests suggestions from our system as well. 

After that, we extract nodes that hold RELATED\_TO relation with our main concepts. After several runs, we decide to deepen our graph traversal technique by employing Dijkstra's shortest path algorithm toward our undirected graphs stored in the graph database. We treat each of the main concepts as a root node and construct separate graphs for these main concepts by traversing through the child nodes until reaching the pre-defined maximum depth which is limited to 2 to limit the size of retrieved nodes. We use these extracted graphs to construct a dictionary with main concepts as keys and child nodes as values. After obtaining the dictionaries for user and project graphs. We initially compare keys in the dictionaries to compare the main concepts for a user story set. 

\begin{figure}[htbp]
    \centering
    \includegraphics[width=\linewidth]{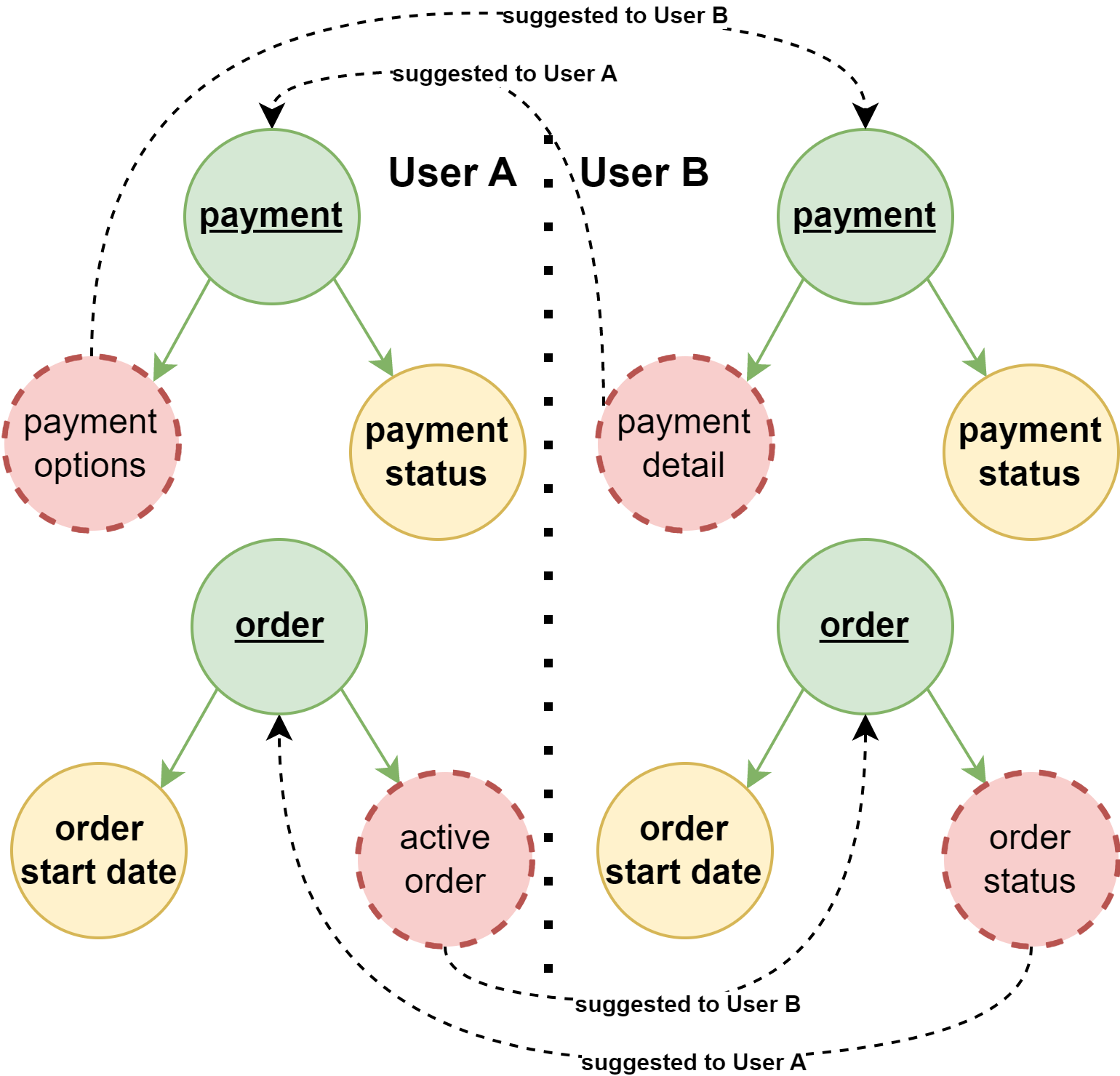}
    \caption{Representation of \textit{close to completeness} suggestion.}
    \label{fig:suggestions-completeness}
\end{figure}

When the user and project graph holds the same main concepts,

\textit{Heuristic 4 - Close to completeness}.
It can be inferred that the user stories that are created by a stakeholder are in alignment with the whole user story set that is constructed by all of the stakeholders for a project. When the main concepts are the same, it is beneficial to compare the concepts that are related to the main concepts. We compare child node values as related concepts and identify missing concepts. We suggest users consider adding these missing child nodes of the main concepts. Fig. \ref{fig:suggestions-completeness} represents the close to completeness heuristic.

When the user and project graph do not hold the same related concepts for any of the main concepts, it can be inferred that a stakeholder mentioned different concepts than other stakeholders. Introducing stakeholders with concepts that they do not include in their user story sets, helps them to easily construct new user stories so that their input enriches the whole user story set. 

\textit{Heuristic 5 - Pop-Zero}.
We consider the main concepts that are extracted from individual stakeholders' input along with their collective input. The difference in the main concepts leads the user story set to form an incomplete solution and also indicates that stakeholders do not share a common way of thinking. 

\heuristics[\linewidth]{-red!75!green!30}{Heuristic 5}{Popular concepts that are less mentioned by the user.}

\begin{figure}[htbp]
    \centering
    \includegraphics[width=\linewidth]{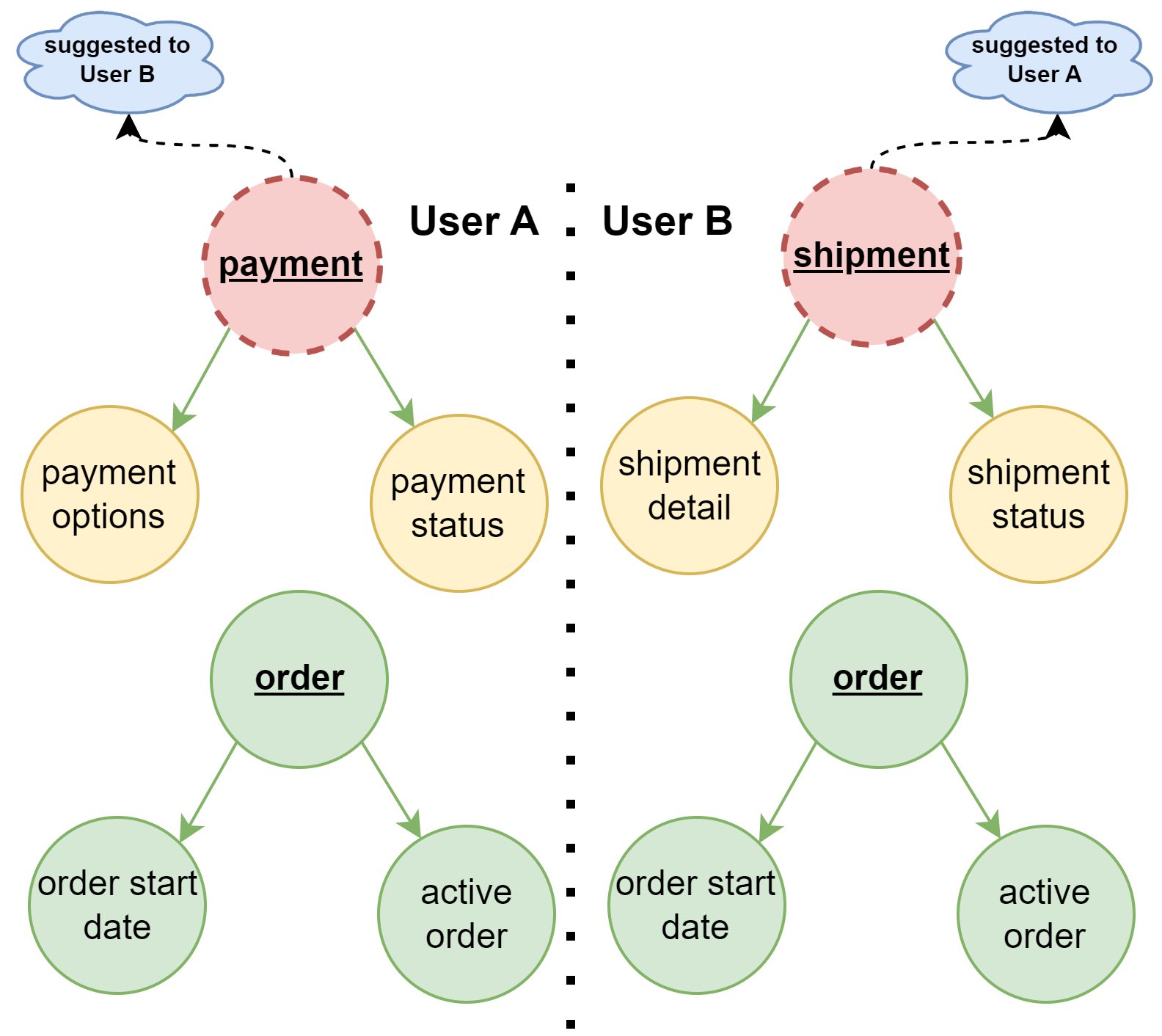}
    \caption{Representation of \textit{heuristic 5}.}
    \label{fig:suggestions-h5}
\end{figure}

Mitigating this issue helps stakeholders to enhance the completeness of the user story set. Therefore, we implement this heuristic that concerns the difference in the main concepts. This heuristic suggests stakeholders introduce new user stories concerning the main concepts that need to be emphasized in detail. Fig. \ref{fig:suggestions-h5} represents the pop-zero heuristic.

\textit{Heuristic 6 - Pop-One}.
In addition to the main concepts, we also take concepts that are related to the main concepts into account. In other words, we benefit from the main concepts and their related concepts. To uncover new concepts contributed by other stakeholders, we extract differences in these related concepts. 


\heuristics[\linewidth]{-red!75!green!30}{Heuristic 6}{Concepts that are not mentioned by the user but are related to the most popular concepts of the project.}

\begin{figure}[htbp]
    \centering
    \includegraphics[width=\linewidth]{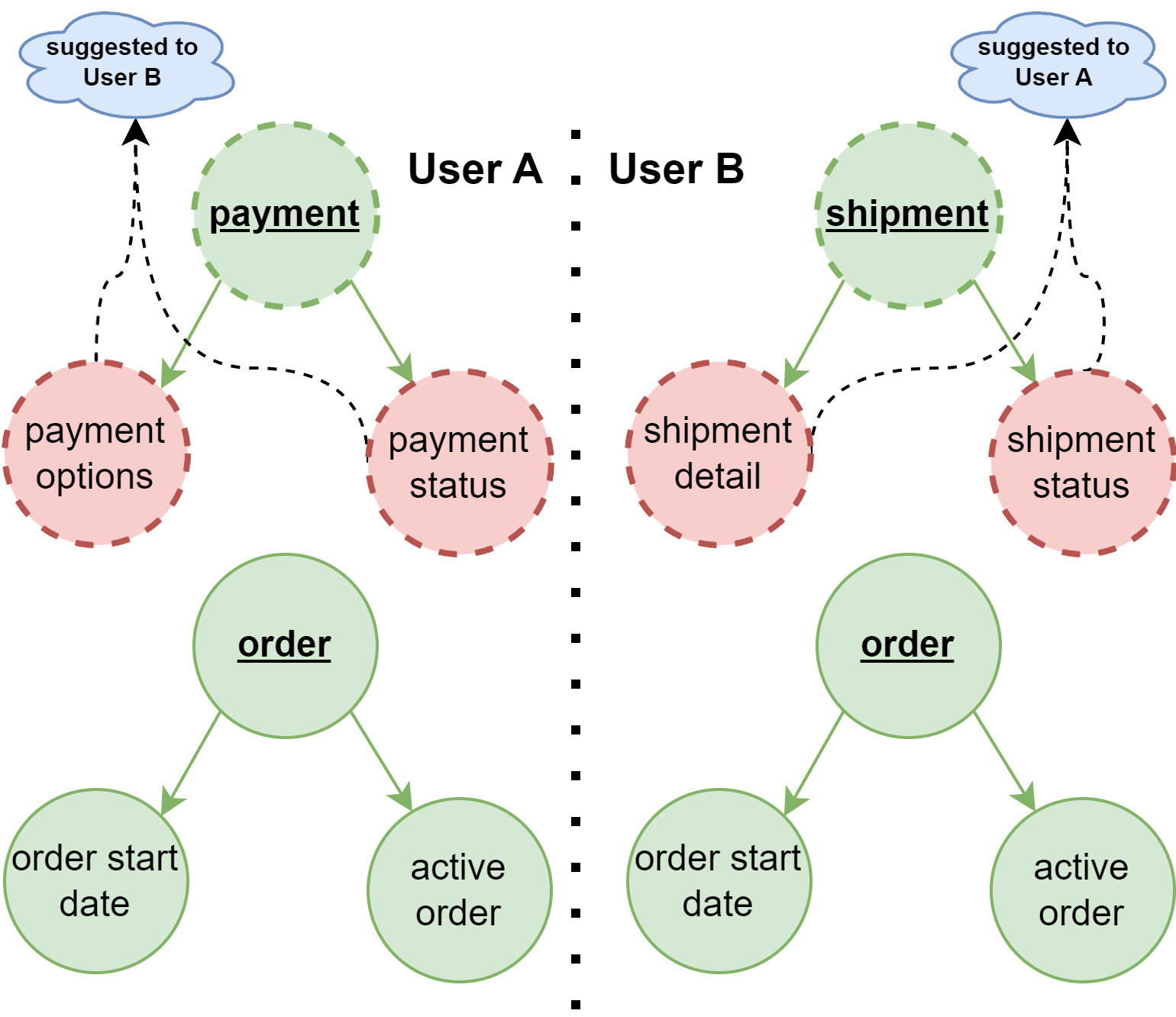}
    \caption{Representation of \textit{heuristic 6}.}
    \label{fig:suggestions-h6}
\end{figure}

This heuristic aims to assist stakeholders by exposing new ideas that are not utilized. By considering these suggestions, stakeholders can be aware of different aspects of a concept that is proposed by their teammates. Fig. \ref{fig:suggestions-h6} represents the pop-one heuristic.


\textit{Heuristic 7 - Pop-Two}.
Besides the main concepts, we consider concepts that have a relation with the main concepts. We extract the main concepts and their related concepts. To uncover new concepts contributed by other stakeholders, we extract differences in these related concepts.

\heuristics[\linewidth]{-red!75!green!30}{Heuristic 7}{Concepts that are only mentioned by the user and are related to the most popular concepts of the project.}

\begin{figure}[htbp]
    \centering
    \includegraphics[width=\linewidth]{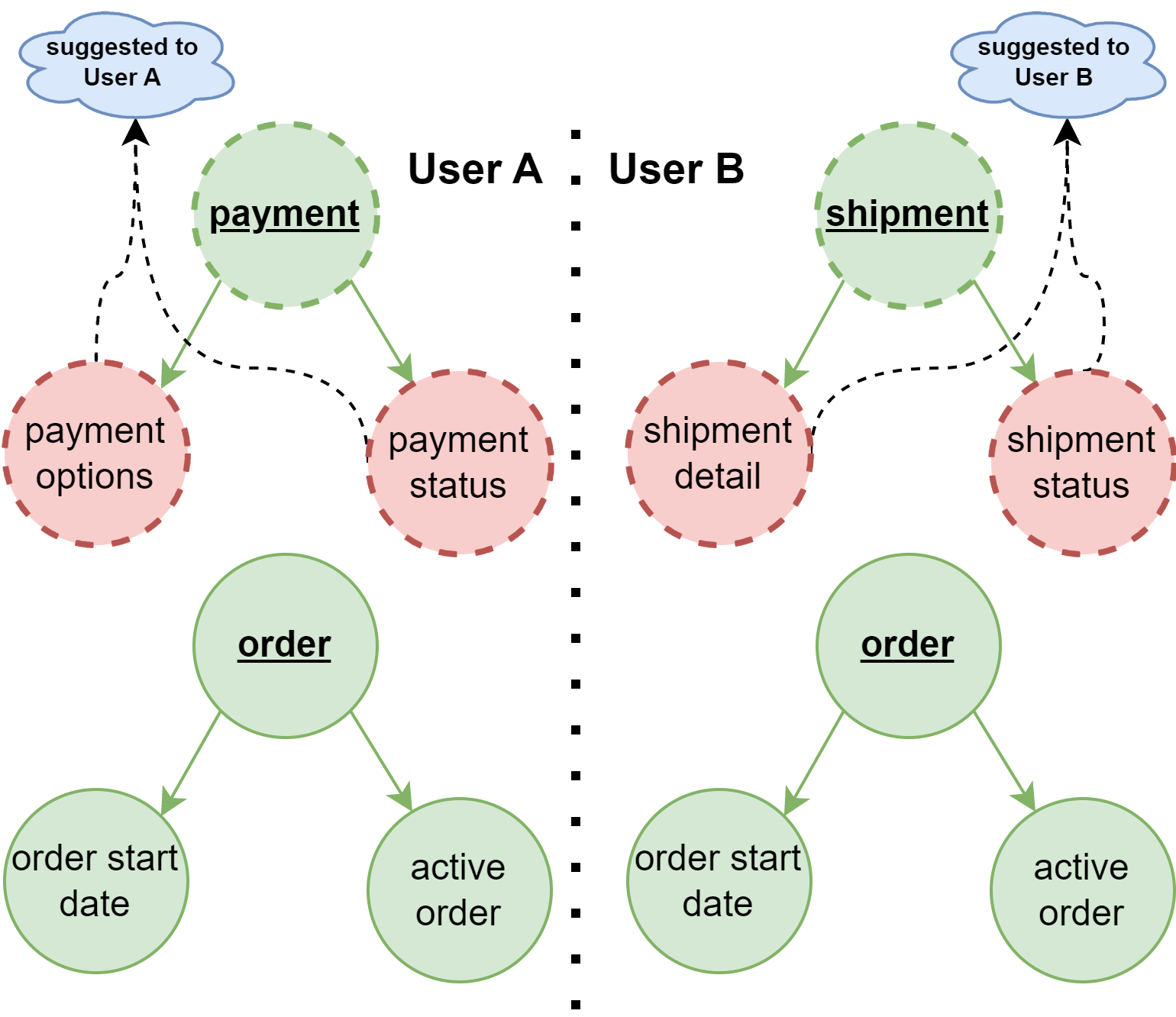}
    \caption{Representation of \textit{heuristic 7}.}
    \label{fig:suggestions-h7}
\end{figure}

This heuristic aims to assist stakeholders in two aspects:
Since any other stakeholder utilizes them in their user story sets,
(i) more user stories about these concepts are necessary for better clarification 
(ii) these concepts can be discarded since they might be out of the scope of that project.
Fig. \ref{fig:suggestions-h7} represents the pop-two heuristic.


\textit{Heuristic 8 - Pop-Three}

When the main concepts of the project are not the same as the main concepts of the stakeholder's input, concepts for some of the stakeholder's input cannot be evaluated. Hence, some of the concepts might be disregarded by the suggestion module. 


\heuristics[\linewidth]{-red!75!green!30}{Heuristic 8}{Concepts that are related to the user's most used concepts are constructed by other members of the project.}

\begin{figure}[htbp]
    \centering
    \includegraphics[width=\linewidth]{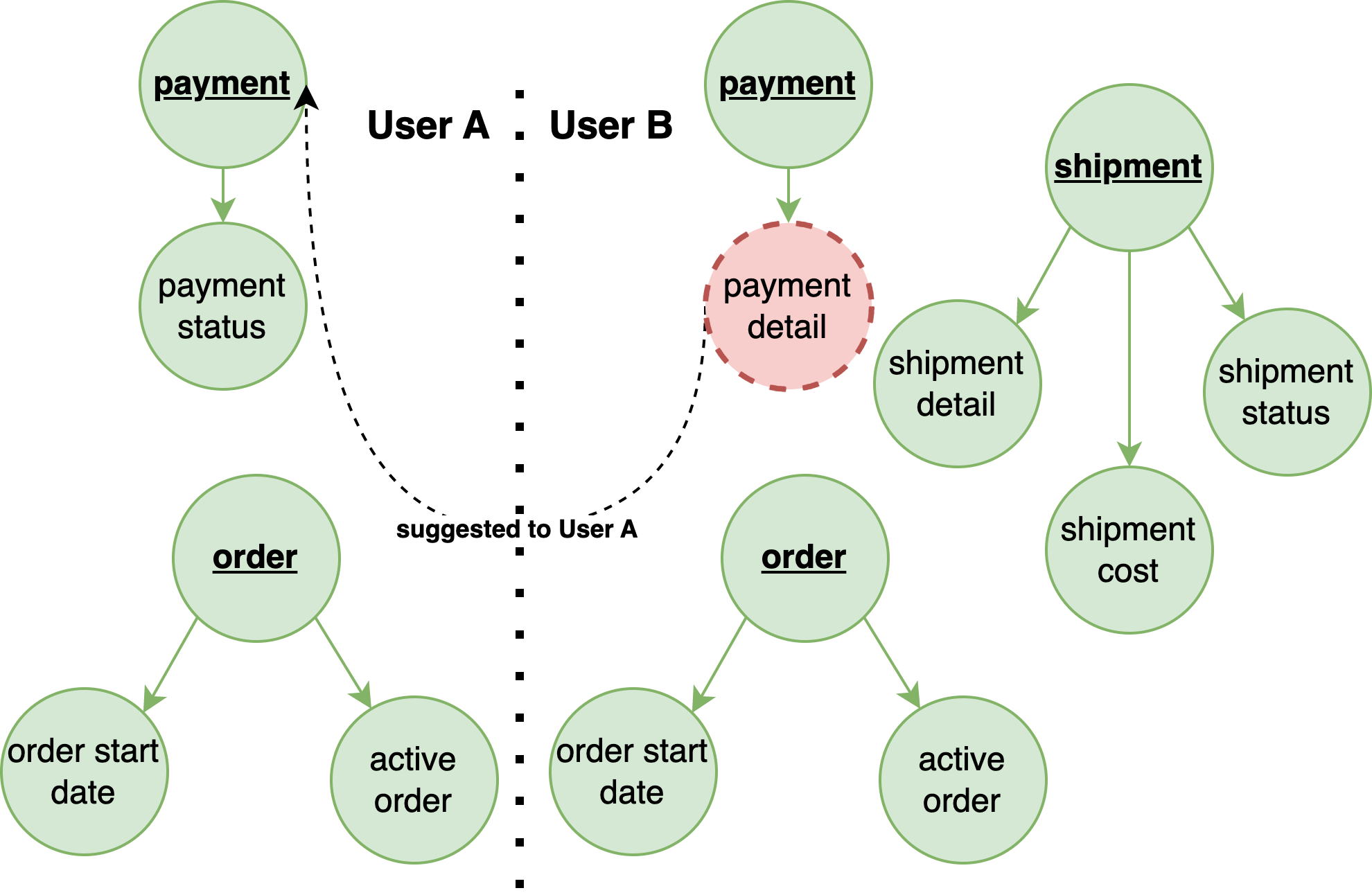}
    \caption{Representation of \textit{heuristic 8}.}
    \label{fig:suggestions-h8}
\end{figure}

To prevent some of the concepts from being disregarded, we compare the stakeholder's main concepts with the other stakeholders' main concepts. By doing so, we enable stakeholders to gather suggestions even if a concept is not widely utilized by all of the members. The idea behind this is to help stakeholders to widen their views and introduce new concepts and user stories. Fig. \ref{fig:suggestions-h8} represents the pop-three heuristic.


\textit{Heuristic 9 - Feeling Lucky?}

When we generate quality suggestions, we extract isolated concepts from knowledge graphs. We generate suggestions about these isolated concepts that are created by other stakeholders for the particular user. This strategy generates the opportunity of collecting other users' opinions about isolated concepts and giving hints to the stakeholders by introducing extremely unfamiliar concepts that might turn into a better perception of the project scope. Fig. \ref{fig:suggestions-h8} represents the feeling lucky heuristic.

\begin{figure}[htbp]
    \centering
    \includegraphics[width=\linewidth]{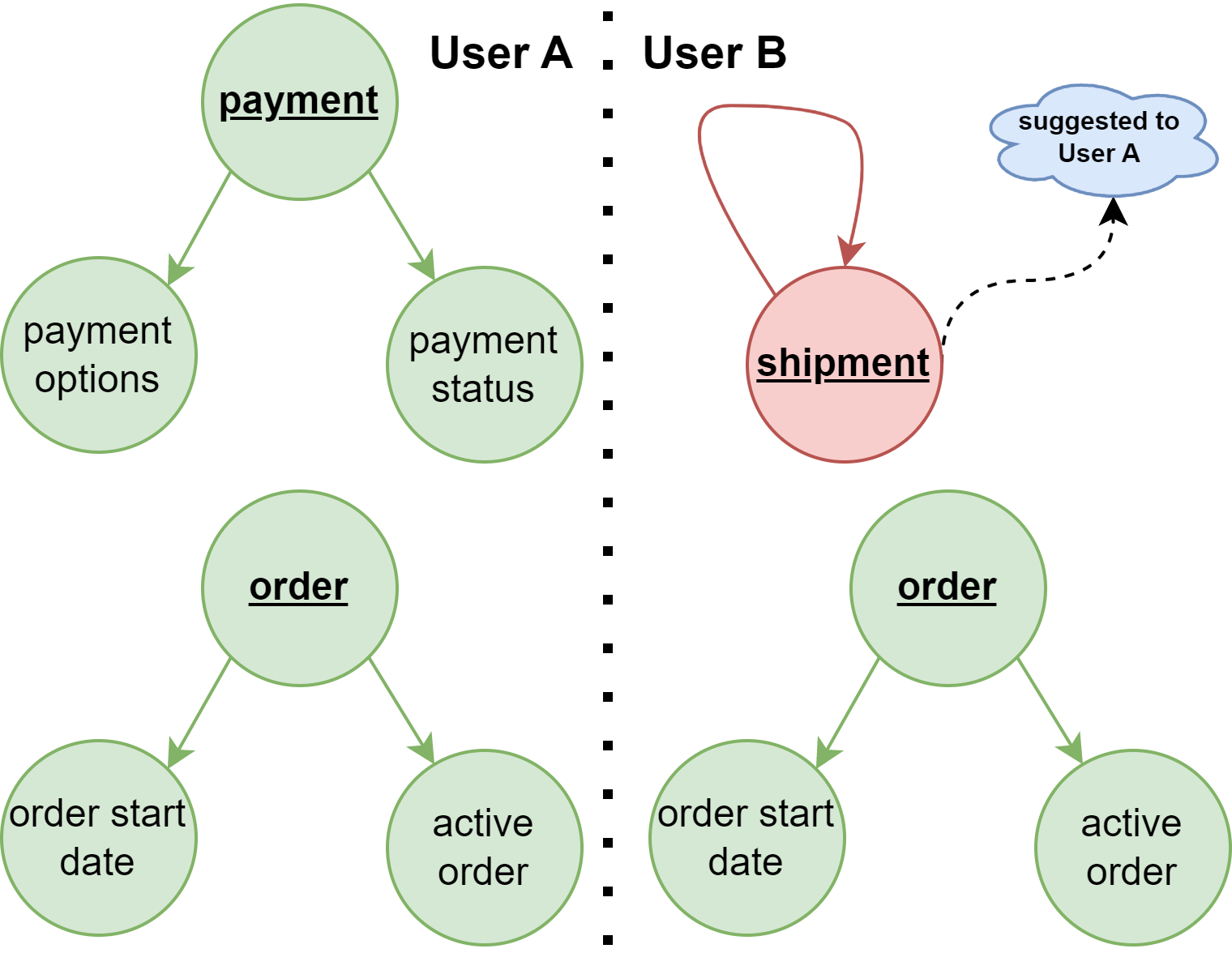}
    \caption{Representation of \textit{heuristic 9}.}
    \label{fig:suggestions-h9}
\end{figure}

\textit{Heuristic 10 - All is well!}
This heuristic is responsible for informing stakeholders that they have successfully achieved a complete solution. We consider both the main and related concepts of the individual stakeholder and the collective input of all stakeholders. When there is not any difference between the main and related concepts at the same time, it can be inferred that the individual stakeholder is aligned with the complete user story set. Therefore, the individual stakeholder is not required to perform any changes.

\textit{Step 7: Suggestion Retrieval}. After the suggestion module generates suggestions using the extracted information. It transfers suggestions to the user interface so that suggestions are displayed to the user.

\section{Implementation}
We use a Linux server that has 1 TB of memory along with an Intel Xeon Gold 6238 CPU with 22 cores that operates at 2.1 GHz. This server runs on Ubuntu 18.04 LTS as the operating system. This server allocates the resources depending on the fair shares principle.

We implement a comprehensive collaborative requirements editor using state-of-the-art technologies that are available for development. We implement our requirements editor using ASP .NET Core. We choose .NET Core for being a lightweight, open-source, and platform-independent framework. We choose commonly used MVC (Model-View-Controller) as an architectural pattern. We implement our chat functionality in our editor using ASP .NET Core SignalR which is an open-source library that provides real-time message broadcasting. 

We employ a Microsoft SQL Server as a database for our editor. By their nature, .NET Core and Microsoft SQL Server work in harmony. To ease our data management processes, we benefit from Entity Framework Core along with Code-First Approach. Therefore, creating our database from scratch takes only seconds, and managing our CRUD operations without writing complex SQL queries provides a flexible implementation phase. 

We implement our web services using Python and Flask. We choose Flask for being a lightweight web framework that can be easily deployed on WSGI HTTP Servers for Unix. We choose gunicorn which is a speedy HTTP Server in our implementation. 

We employ neo4j as our graph database for being open-source and providing high performance. We use neo4j Bolt Driver and Cypher Query Language to extract information from the graph databases. 

We containerize the different parts of our system using Docker to increase maintainability and management. We use official Docker images for Microsoft SQL Server and neo4j. We manually create Docker containers from scratch by constructing different Dockerfiles for different modules of the system. As a result, we use 4 different docker containers that Docker Compose orchestrates.

\section{Evaluation}
\label{section:evaluation}

This section presents the design, execution, and results of the conducted experiment. Our goal is to create an effective method compared to a standard collaborative writing tool in terms of constructing user stories. 

We hypothesize that, H0 = "the level of completeness for a user story set constructed with \sys{} is lower or equal to a user story set constructed with an average collaborative text editor". The one-tail alternative hypothesis is H1 = "the level of completeness for a user story set constructed with \sys{} is greater than a user story set constructed with an average collaborative text editor".

Hence, we provide an efficient way of constructing a coherent user story set for requirements engineers. Therefore, we evaluate our system by conducting experiments with human participants to measure the impact of \sys{} on the task of constructing user story sets.


\subsection{Experimental design}
To evaluate \sys{}, we designed an experiment that involves participant engagement with the system. Participants are randomly assigned to different setups among two available setups. The experiment requires participants to write as many user stories as possible within a period. 

\begin{figure}[htbp]
    \centering
    \includegraphics[width=0.5\linewidth]{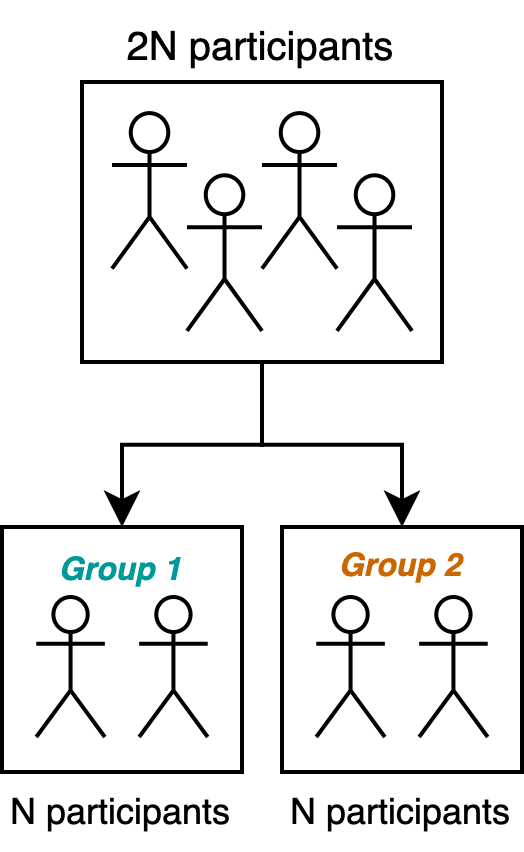}
    \caption{Participant distribution}
    \label{fig:stickmen}
\end{figure}

\begin{figure}[htbp]
    \centering
    \includegraphics[width=\linewidth]{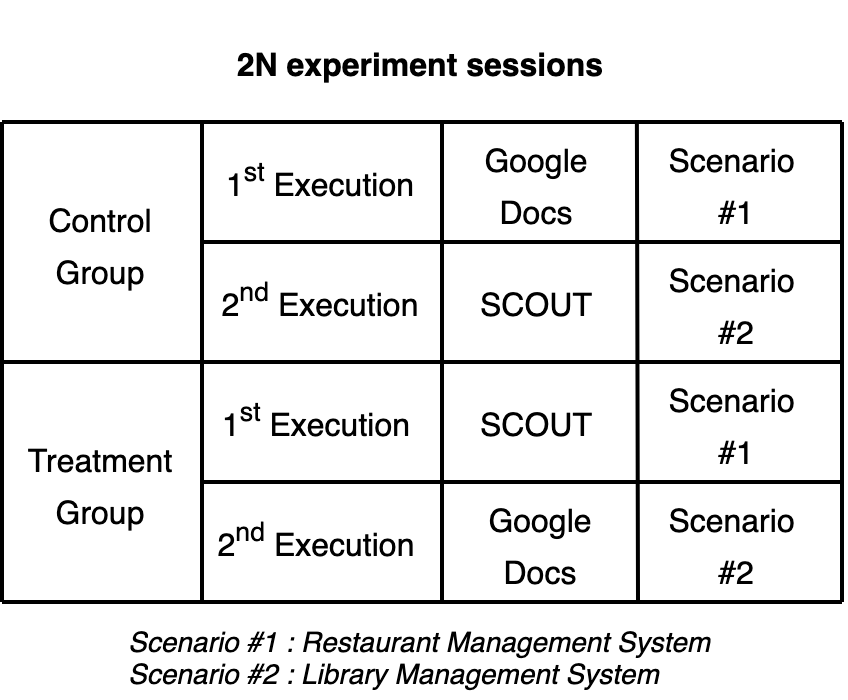}
    \caption{Experimental design}
    \label{fig:sessions}
\end{figure}

\subsection{Procedure}

We plan to conduct online experiment sessions with the participants via an online conference tool due to Covid-19 restrictions. We inform potential participants about our approach via a written document that contains a brief explanation about \sys{} along with the experimental setup. We also inform the potential participants about the average duration of the experiment so that they can arrange their available periods, if possible. After that, we invite them to participate in our experiment voluntarily. We try to gather participants who mostly have similar backgrounds. However, we also conduct experiments with participants that work outside of the IT industry and undergraduate students to increase the coverage of our experiment. We aim to widen the application area of \sys{} as well as encourage other parties to employ user stories to support their business activities. 

We arrange available time slots with those who volunteered to be a participant in our experiment. Before the experiment, we get each participant's consent to participate in the experiment. We also collect demographic information from participants to recognize our target group. 

After completing the pre-experimental steps, the experimenter makes a presentation that includes the aim of the experiment, experimental setup, experimentation timeline, and usage of the \sys{} along with the logic behind the different suggestion types. Participants were asked to complete the experiment on their own. We provide chat functionality in \sys{} to provide a communication channel between participants. Additionally, they do not require any additional tool or application to communicate with each other. The goal is to avoid any participant bias caused by the help of others. The experimenter is also present in the online meeting room throughout the experimentation to observe participants. 

\subsection{Materials}

\textit{Scenarios.}
We conduct our experiments using two different scenarios to mitigate the effect of the learning curve. We choose the development of \textit{restaurant and library} management systems as our scenarios since they do not require any advanced expertise. Additionally, these scenarios are widely used examples for educational purposes thus eliminating the risk of one group dominating another one. The provided scenarios are publicly available.

\textit{Pre-experimental survey.}
We collect general demographic information from the participants (i.e age, gender, occupation, and duration of professional experience). We also asked participants about their source of user story knowledge, whether they only employ user stories for educational purposes or they also employ user stories in their professional life, and their level of confidence and understanding while employing user stories.

\begin{table*}[ht!]
\caption{Questions for pre-experimental survey}
\label{pre-exmerimental-table}
\resizebox{\textwidth}{!}{%
\begin{tabular}{@{}|l|l|l|l|l|l|@{}}
\hline
\rowcolor[HTML]{DAE8FC} 
Q1 & \multicolumn{5}{l|}{\cellcolor[HTML]{DAE8FC}How many years of professional experience working in the IT industry do you have?} \\\hline
   & 0             & 1-3                 & 4-6                 & \multicolumn{2}{l|}{More than 6}                                \\\hline
\rowcolor[HTML]{DAE8FC} 
Q2 &
  \multicolumn{5}{l|}{\cellcolor[HTML]{DAE8FC}\begin{tabular}[c]{@{}l@{}}How many times that you participate in constructing user story sets for a  software development\\ project? (class projects and assignments)\end{tabular}} \\\hline
   & 0             & 1-3                 & 4-6                 & \multicolumn{2}{l|}{More than 6}                                \\\hline
\rowcolor[HTML]{DAE8FC} 
Q3 &
  \multicolumn{5}{l|}{\cellcolor[HTML]{DAE8FC}\begin{tabular}[c]{@{}l@{}}How many times that you participate in constructing user story sets for a  software development \\ project? (conducted in an agile development cycle)\end{tabular}} \\\hline
   & 0             & 1-3                 & 4-6                 & \multicolumn{2}{l|}{More than 6}                                \\\hline
\rowcolor[HTML]{DAE8FC} 
Q4 & \multicolumn{5}{l|}{\cellcolor[HTML]{DAE8FC}Are you actively using user story sets in your daily workflow?}                 \\\hline
   & 1             & 2                   & 3                   & 4                       & 5                                    \\\hline
\rowcolor[HTML]{DAE8FC} 
Q5 &
  \multicolumn{5}{l|}{\cellcolor[HTML]{DAE8FC}\begin{tabular}[c]{@{}l@{}}Which kind of training do you have to learn how to adapt user stories into software \\ development phases?\end{tabular}} \\\hline
   & Self study    & University course   & Company trainings   & Industrial experience   & \makecell{I do not have \\ any training on this}   \\\hline
\rowcolor[HTML]{DAE8FC} 
Q6 & \multicolumn{5}{l|}{\cellcolor[HTML]{DAE8FC}My level of understanding on the theory of user stories}                        \\\hline
   & 1             & 2                   & 3                   & 4                       & 5                                    \\\hline
\rowcolor[HTML]{DAE8FC} 
Q7 & \multicolumn{5}{l|}{\cellcolor[HTML]{DAE8FC}I feel confident while employing user story sets in my projects.}               \\
   & 1             & 2                   & 3                   & 4                       & 5        \\\hline                           
\end{tabular}%
}
\end{table*}

\begin{table*}[ht!]
\caption{Questions for post-experimental survey}
\label{post-exmerimental-table}
\resizebox{\textwidth}{!}{%
\begin{tabular}{@{}|l|l|l|l|l|@{}}
\hline
\rowcolor[HTML]{B9FCB8} 
\makecell{Strongly Disagree \\ 1} & \makecell{Slightly Disagree \\ 2} & \makecell{Neither Agree Nor Disagree \\ 3} & \makecell{Slightly Agree \\ 4} & \makecell{Strongly Agree \\ 5} \\\hline
\multicolumn{5}{|l|}{This tool helped me to improve the quality of the user stories I write.}                    \\\hline
\multicolumn{5}{|l|}{This tool helped me to write more user stories.}           \\\hline
\multicolumn{5}{|l|}{The user interface is user-friendly.}                      \\\hline
\multicolumn{5}{|l|}{The suggestions provided by the tool were useful.}         \\\hline
\multicolumn{5}{|l|}{If the tool is available, I would use it for my projects.} \\\hline
\end{tabular}%
}
\end{table*}

\textit{Post-experimental survey.}
We collect participants' opinions about \sys{} as well as their feedback.

\paragraph{Questionnaire on Effectiveness.} We ask participants how effective \sys{} is in terms of increasing user story quality and increasing the number of user stories. We also collect their opinion on whether \sys{} provides a user-friendly interface.
\paragraph{Questionnaire on Suggestion Quality.} We asked participants how useful the suggestions that are generated by our system are. We also collect the most and least favorite types of suggestions to determine the level of acceptance of our suggestion strategies.
\paragraph{Questionnaire on Expected Futures.} We asked participants about the positive and negative aspects of \sys{}. The responses are collected as free texts and examined by the authors. By asking these questions, we aim to keep our strong aspects whilst improving the weak aspects of our system.

\subsection{Study Execution}
We conduct our experiment in 70 minutes. We set 30 minutes for each experimental run and additional 10 minutes for the surveys. We split the participants into two groups control and treatment groups to conduct different steps of our evaluation. We use \sys{} along with a collaborative writing editor for measuring the impact of \sys{}. To keep experiment scenarios in their eyesight we put the brief scenario description on the top of the collaborative writing editor and our homepage. 

For the first run, participants in the control group will be asked to create user stories for the first scenario (S1) by using a collaborative writing editor as a group. Meanwhile, participants in the treatment group will be introduced to \sys{} and asked to use \sys{} to create user stories for the second scenario (S2). After the first run is completed, participants in the control group are introduced to \sys{} for the first time and asked to use \sys{} to create user stories for the second scenario (S2). Meanwhile, participants in the treatment group will be introduced to a collaborative writing editor and asked to use this collaborative writing editor to create user stories for the first scenario (S1). During the experiments with \sys{}, the participants are kindly reminded that \sys{} has a chat functionality for communicating with other members of their group and \sys{} can provide suggestions when requested. After the second run, we will be able to gather data for further calculations to create quantitative feedback from the participants. To gather qualitative results, participants are asked to fill out our post-experiment survey.

\subsection{Results}
%
%
%
%
%
%

\subsubsection{Demographics} 
\begin{figure}[htp]
\centering
  \includegraphics[width=\linewidth]{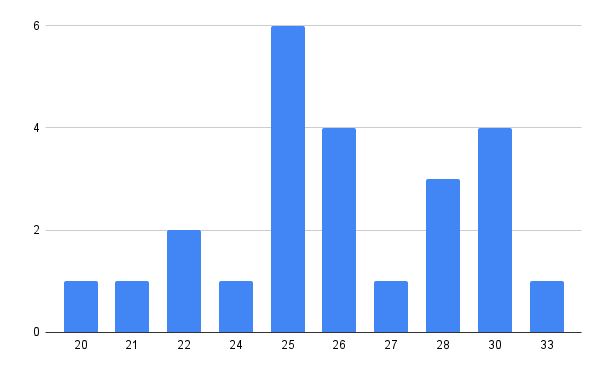}
  \caption{Age distribution of participants} 
  \label{fig:ages}
\end{figure}

We experiment with a participant group of 24 people who already know user stories. The participant group contains individuals between the ages of 20 and 33. The average age of the participants is 26.1. The age distribution of participants can be seen in Fig. \ref{fig:ages}

\begin{figure}[htp]
\centering
  \includegraphics[width=\linewidth]{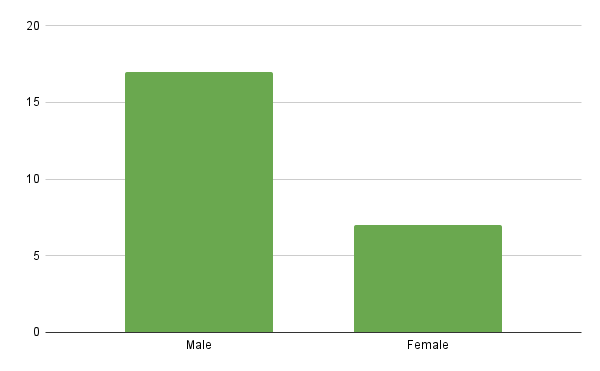}
  \caption{Gender distribution of participants} 
  \label{fig:gender}
\end{figure}

The participant group has a gender ratio of 70.83\% for male participants and 29.17\% for female participants. The gender distribution of participants can be seen in Fig. \ref{fig:gender}

We collect the participants' occupations via a demographics survey. 11 of the participants (45.83\%) are occupied as software developers. 
1 of the participants occupied as a software developer also stated that he/she is an undergraduate student. 
2 of the participants occupied as software developers also stated that he/she is a graduate student. 
7 of the participants (29.16\%) occupied as data engineers/scientists. 
1 of the participants occupied as a data engineer/scientist also stated that he/she is an undergraduate student.
2 of the participants (8.33\%) are occupied as electronics engineers/engineers. 1 of the participants (4.16\%) is occupied as a risk analyst. 2 of the participants (8.33\%) are occupied as undergraduate students. 1 of the participants (4.16\%) is occupied as a graduate student.

\begin{figure}[htp]
\centering
  \includegraphics[width=\linewidth]{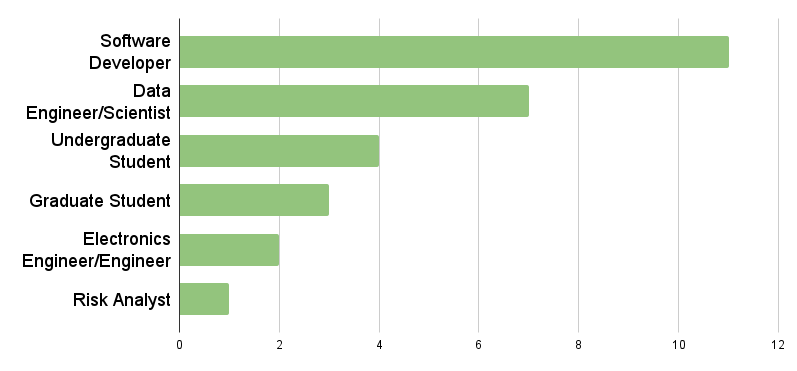}
  \caption{Occupation of participants} 
  \label{fig:occupation}
\end{figure}

\begin{figure}[htbp]
\centering
  \includegraphics[width=\linewidth]{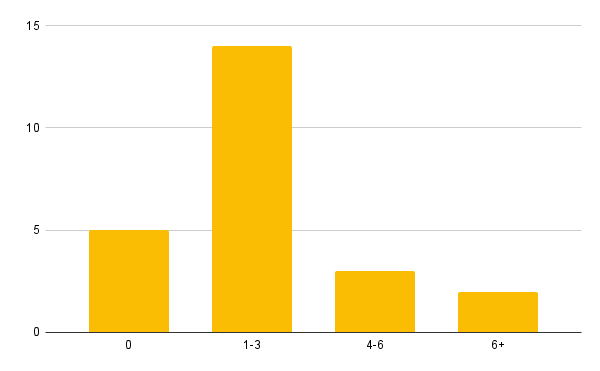}
  \caption{IT industry experience levels of participants} 
  \label{fig:experience}
\end{figure}

We collect the professional experience that participants have in the IT industry. Results show that 5 of the participants have zero experience in the IT industry. 2 of them are undergraduate students, and 3 of them work outside of the IT industry. Additionally, 14 participants claim that they have 1-3 years of IT industry experience, 3 participants claim that they have 4-6 years of IT industry experience and 2 participants claim that they have more than 6 years of IT industry experience. Fig. \ref{fig:experience} shows the level of experience that participants have.

We also asked different questions to retrieve information about their prior experience with the user stories. 

First, we check for their prior experience with user stories in their class projects or assignments. 7 of the participants claim that they did not participate in any class project or assignments that require constructing user stories. 14 of the participants claim that they participated in 1-3 projects. 2 participants claim that they participated in 4-6 projects and 1 participant claim that he/she participated in more than 6 projects. 

\begin{figure}[htbp]
\centering
  \includegraphics[width=\linewidth]{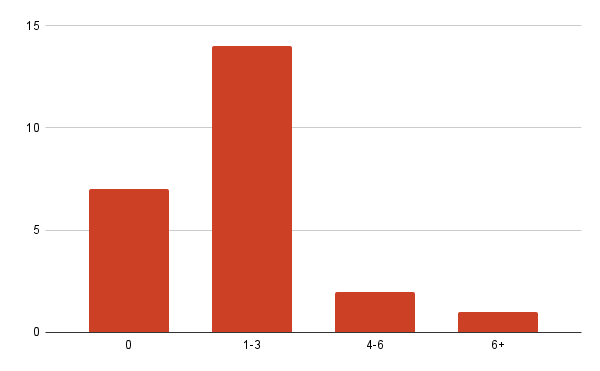}
  \caption{How many times that you participate in constructing user story sets for a software development project? (class projects and assignments)} 
  \label{fig:classproj}
\end{figure}

Second, we check for their prior experience with user stories in an agile development cycle. 12 of the participants claim that they did not participate in any agile development projects or they did not use user stories in an agile development cycle. 7 of the participants claim that they participated in 1-3 projects. 4 of the participants claim that they participated in 4-6 projects and 1 participant claim that he/she participated in more than 6 projects.

\begin{figure}[htbp]
\centering
  \includegraphics[width=\linewidth]{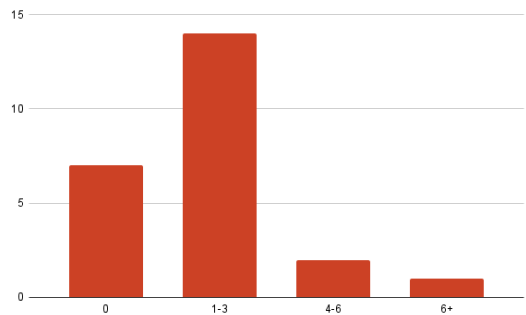}
  \caption{How many times that you participate in constructing user story sets for a software development project? (conducted in an agile development cycle)} 
  \label{fig:agiledev}
\end{figure}

We also ask the participants whether they use user stories in their daily workflow. The question is asked in a form of a 5-point Likert scale. 10 of the participants are marked as 3 points or higher and 14 of the participants are marked as 1 or 2. Even though half of the users construct user stories in an agile development cycle, agile development practices seem not to opt for daily workflow for most of the practitioners.

\begin{figure}[htbp]
\centering
  \includegraphics[width=\linewidth]{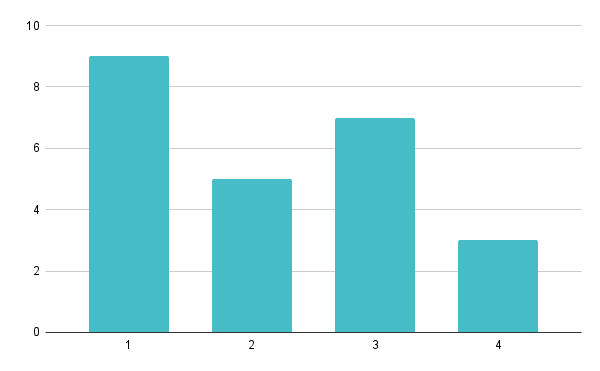}
  \caption{Are you actively using user story sets in your daily workflow?} 
  \label{fig:activeusing}
\end{figure}

We also asked the participants about their level of theoretical understanding of user stories. The questions are asked in a form of a 5-point Likert scale. 17 of the participants were marked as 3 points or higher and 7 of the participants were marked as 1 or 2. It shows that majority of participants have an adequate understanding of the theory of user stories.

\begin{figure}[htbp]
\centering
  \includegraphics[width=\linewidth]{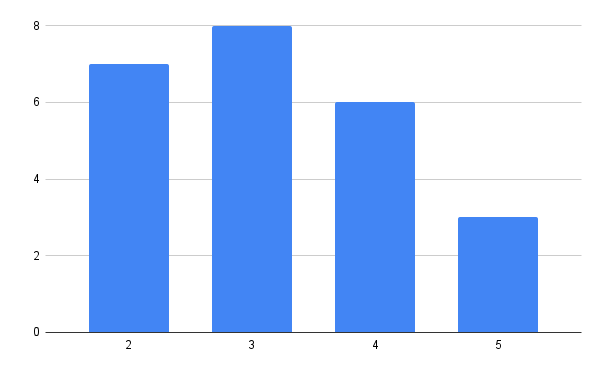}
  \caption{Level of understanding on the theory of user stories} 
  \label{fig:understanding}
\end{figure}

We also asked the participants whether they are confident while employing user stories in their projects. The questions are asked in a form of a 5-point Likert scale. 17 of the participants were marked as 3 points or higher and 7 of the participants were marked as 1 or 2. It shows that majority of participants are comfortable with user stories.

\begin{figure}[htbp]
\centering
  \includegraphics[width=\linewidth]{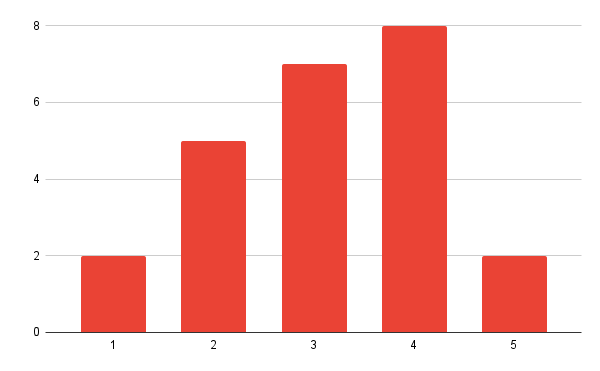}
  \caption{Level of confidence while employing user story sets} 
  \label{fig:confident}
\end{figure}

We also asked participants for their source of knowledge about user stories. 10 of the participants claim that university courses, 4 of the participants claim that industrial experience, 16 of the participants claim that self-study, 9 of the participants claim that company training is their source of user story knowledge, and 3 of the participants claim that they do not have any training on user stories.

\begin{figure}[htbp]
\centering
  \includegraphics[width=\linewidth]{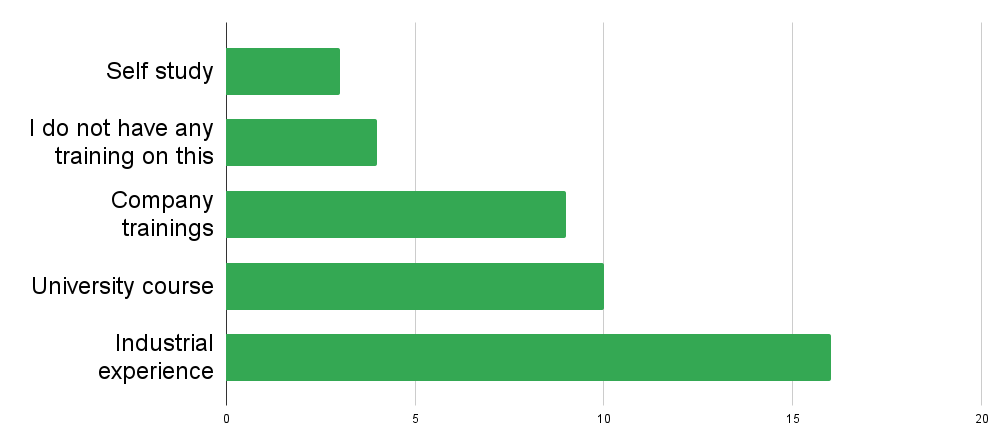}
  \caption{Source of knowledge about user stories} 
  \label{fig:knowledge}
\end{figure}

In general, demographic survey results indicate that our participants have adequate theoretical knowledge and practical experience with user stories.

\subsubsection{Results}

\paragraph{\textbf{RQ3:} How effective is our system in terms of increasing the completeness of the user story set?}
During our experiment, participants construct 330 user stories using a collaborative writing editor and 448 user stories using \sys{}. Participants manage to construct 35.76\% more user stories by using \sys{}. Additionally, we examine each user story that is constructed by participants. We observe an average number of 27.5 user stories per project in the collaborative writing editor. On the other hand, We observe an average number of 37.5 user stories per project with \sys{}. 

We also examine the standard deviation of the number of user stories among projects. We observe that the standard deviation of the number of user stories differs significantly. We observe 10.77 for the collaborative writing editor and 9.59 for the \sys{}. To examine this difference in detail, we check whether the samples come from a normal distribution. Results show that number of user stories that are created via using the \sys{} is normally distributed. On the other hand, the number of user stories that are created using a collaborative writing editor is not normally distributed. To check the statistical significance between two independent sample sets, we applied Mann-Whitney U Test since the size of the sample set is relatively small (n = 12 and n < 20). As a result of the Mann-Whitney U test, we obtain a p-value of 0.015 which is smaller than the alpha value of 0.05. Therefore, these results are independent of each other. The difference in variance can also be explained by the concept of writer's block. The lack of assistance to the participants when they need additional ideas causes the number of constructed user stories to remain low. Some of the groups might not face any problem when it comes to constructing user stories continuously however participants with less prior experience with user stories suffer from this problem.

We also examine the graph data. As an initial assumption, we believe that we encourage users to add more definitions to the phrases that they used. When we consider all of the noun phrases as a tree, including different properties of a term yields more branches added to a single node. To examine this assumption, we convert the stored graphs into trees and apply the breadth-first search starting from the node that has the lowest degree. During the breadth-first search, we collect the number of edges that are included. We observe that graphs that are generated whilst using \sys{} have 32.46\% more edges compared to the user stories that are generated by using a collaborative writing editor. We also observe a slight increase (6.55\%) in average node connectivity with \sys{}.

Our null hypothesis relies on the definition that the level of completeness for a user story set constructed with an average collaborative text editor is lower or equal to a user story set constructed with \sys{}. However, the results indicate that \sys{} performs significantly better compared to Google Docs. Therefore, we reject the null hypothesis (H0).

\paragraph{\textbf{RQ4:} To what extent do the participants implement the suggestions produced by \sys{}?}
We examine the applied suggestions by participants in two groups as quality and completeness suggestions. We generate 453 distinct quality suggestions and participants choose to apply (33.11\%) of them. We also generate 991 completeness suggestions and participants choose to apply (70.74\%) of them. We investigate deeply for each suggestion type.

When we look at the number of generated and applied quality suggestions, CRUD suggestions are the least preferred suggestions among others. 57 of the 271 (21.03\%) CRUD suggestions are applied by participants. On the other hand, 5 of the 7 (71.43\%) atomic suggestions are applied by participants. A higher portion of applied atomic suggestions yields better quality in individual user stories. Additionally, 88 of 175 (50.29\%) isolated suggestions are applied by participants. By generating isolated suggestions, we point out the concepts that need clarification or better explanation so that these concepts are explained in depth. A higher portion of applied isolated suggestions prevents user stories from being ambiguous. The results indicate that our quality suggestion heuristics are accepted by participants and used to increase the individual user story quality.

When we look at the number of generated and applied completeness suggestions, we observe promising results. 

76 of 116 (65.52\%) close to completeness suggestions are applied by the participants. When participants do not face the differences in main concepts, they easily put effort to eliminate the incompleteness in the user story set. 

62 of 83 (74.70\%) pop-zero suggestions are applied by the participants. 
218 of 369 (59.08\%) pop-one suggestions are applied by the participants. 
90 of 109 (82.57\%) pop-two suggestions are applied by the participants. 
216 of 256 (84.38\%) pop-three suggestions are applied by the participants. 

Results show that participants also highly adopt the suggestions that are generated due to the difference in main concepts compared to the whole user story set. The idea behind these suggestions is to enlighten different viewpoints for users to widen their way of thinking.

39 of 58 Feeling Lucky? (67.24\%) suggestions are applied by participants. 
We introduce participants to isolated concepts that are created by other users. They use isolated concepts of other users as a reference point to construct user stories that clarify these terms. Therefore, the higher portion of application of these suggestions directly reduces the level of incompleteness in the user story set. 

\paragraph{\textbf{RQ5:} What is the perceived usability level?}
After the experiment, we conduct a post-experimental survey to collect qualitative results from the participants. We did not obligate participants to attend any part of the experiment. They conduct this experiment voluntarily. Yet, all of them participate in the post-experimental survey and evaluate the usability and suggestion performance of \sys{}. 

First, we asked participants whether \sys{} improve the quality of the user stories that they wrote during the experiment. This question is asked in a form of a 5-point Likert scale. 3 of them marked 3 points, 14 of them marked 4 points and 7 of them marked 5 points. \sys{} achieves an average point of 4.17 out of 5. We can infer that participants believe that they write better user stories with \sys{}.

We also asked participants whether \sys{} help them to increase the number of user stories. This question is asked in a form of a 5-point Likert scale. 2 of the participants marked 3 points, 6 of them marked 4 points and 16 of them marked 5 points. \sys{} achieves an average point of 4.58 out of 5. We can infer that participants benefit from \sys{} to increase the number of user stories when needed. Quantitative results also show a significantly higher average number of user stories and lower variance in the number of user stories per project. Since participants are assigned randomly to the groups, some of the participants may have more prior experience with the user stories resulting in the excessive difference in the number of user stories among projects. When \sys{} is used, all of the participants have the chance to get support when they need it. Therefore, we can infer that \sys{} helps users to generate more user stories.

We also asked participants whether the user interface of \sys{} is user-friendly. This question is asked in a form of a 5-point Likert scale. 5 of the participants marked 3 points, 12 of the participants marked 4 points and 7 of them marked 5 points. \sys{} achieves an average point of 4.08 out of 5. We can infer that users easily get used to \sys{} without any distractions. While creating the interface of \sys{}, we try to make it as simple as we can. We use simple components, format instructions, and explanations for suggestions along with functionalities that increase collaboration between users such as a chat panel.

We also asked participants whether the suggestions generated by \sys{} are useful. This question is asked in a form of a 5-point Likert scale. 1 of the participants marked 3 points, 16 of the participants marked 4 points and 7 of them marked 5 points. \sys{} achieves an average point of 4.25 out of 5. We can infer that most of the users benefit from the suggestions. Suggestions are either directly applied to increase quality and completeness or create better ideas in the participants' minds and lead them to generate more comprehensive user stories. As aligned with the quantitative results, a great portion of the generated suggestions is applied or suggested points corrected by the participants. 

We also asked participants whether they would use \sys{} for their projects. This question is asked in a form of a 5-point Likert scale. 1 of the participants marked 2 points, 3 of them marked 3 points, 12 of them marked 4 points and 8 of them marked 5 points. \sys{} achieves an average point of 4.13 out of 5. We can infer that most of the users are willing to use \sys{}. Even though being in the early phases of development, \sys{} achieved positive results in terms of acceptance of the approach.

We asked participants to state their favorite features of \sys{}. We collect 24 free text answers from the participants. We categorized them into 2 different groups suggestions and editors. 21 of the answers indicate that suggestions are by far the most favorite feature of \sys{} followed by 4 answers for the editor. Our main goal with \sys{} is to support stakeholders by generating suggestions and both quantitative and qualitative results reveal that our technique is opted for by participants.

We asked participants to state their most and least useful types of suggestions. We collect 24 answers for each type of evaluation from the participants that contain main suggestion groups (quality and completeness) and sub-suggestion groups (isolated, atomic, CRUD, etc.). 3 of the participants decide not to share their least type of suggestions. We categorized the answers into two groups quality suggestions, completeness suggestions, none, and both. 14 of the answers state that one or more of the quality suggestions as their favorite type of suggestions. On the other hand, 14 of the answers state that one or more of the quality suggestions was their least type of suggestion. 11 of the answers state that one or more of the completeness suggestions was their favorite type of suggestion. On the other hand, 7 of the answers state that one or more of the completeness suggestions was their least type of suggestion. 6 of the least favorited suggestions come from feeling lucky types of suggestions. It seems that users have a hard time using that type of suggestion. Combining the results of the applied suggestions and participants' favorite type of suggestion. Completeness suggestions are widely embraced by the participants. 

It can be inferred that completeness suggestions are preferred by the participants. This result is aligned with the quantitative results because a higher portion of the completeness suggestions is applied by the participants during the experiment.

We also collect participants' positive and negative feedback along with their suggestions to make \sys{} better. we collect 24 positive feedback from the participants that refer to \sys{} as helpful.
We collect 15 negative feedbacks to be considered to improve \sys{} with 24 participants. The remainder of the participants does not give any feedback. We categorize the suggestions into 3 different groups suggestions, performance, and editor. 9 of the participants give feedback for improvements in suggestions generated by \sys{}. 1 of the participants was concerned about the non-functional requirement of the performance of the \sys{}. 5 of the participants suggest improvements to increase the usability of \sys{}. We evaluate these suggestions and concerns to form our future work.

\section{Conclusion}
\label{section:conclusion}
This paper introduces a collaborative requirements editor instrumented with an NLP-powered suggestion functionality. As a component of the NLP pipeline, our study uses a pre-trained deep language model (BERT), which has enormous potential for requirements engineering activities. We provide the publicly accessible source code together with information on how the entire system was designed and how it was implemented. Additionally, we contribute to the community by sharing the user stories we gather during the experiments to build a dataset for consequent user story research. We evaluate \sys{} by comparing human performance while using a collaborative text editor with \sys{}. By calculating several metrics and conducting statistical tests, we quantitatively evaluated \sys{}. The quantitative results suggest that by utilizing \sys{}, stakeholders can create user story sets that are noticeably more complete. A statistical significance test supports this inference. Additionally, because \sys{} offers stakeholders on-demand suggestions while they are creating user stories, it cuts down on the time and iterations necessary for stakeholders to review the user story sets. The findings also demonstrate how eagerly the participants adopted the provided suggestions. The qualitative results suggest that participants are satisfied with \sys{} in a variety of ways. Participants had no problems utilizing \sys{} since they are highly satisfied with the user interface. Additionally, a significant number of users claimed that \sys{} is helpful by offering new ideas to improve the completeness and quality of the user story sets. The majority of the participants also state that if \sys{} is accessible, they will employ it. Overall results suggest that requirements engineering experts would utilize \sys{}. Companies that use agile development principles may simply adopt our technology to their daily workflows since industry awareness of the advantages of requirements engineering practices is continually growing. Comparing this adaption to manual processes may result in lower costs and increased efficiency. We manage to achieve good results with \sys{}. We plan to develop the following concerns as future work to carry \sys{} out.

\textit{Prevent showing the disliked suggestions}. Users can indicate that they dislike some suggestions by clicking the crossed-eye icon in the UI. We collect this feedback to decide on further steps of our implementation to improve the quality of our suggestions. For now, we do not hide these disliked suggestions but we implement this functionality in future work.

\textit{Improving the NLP pipeline}. In this work, we benefit from the input of the stakeholders in a collaborative project. We aim to expand \sys{} by generating suggestions that contain concepts that stakeholders have never seen before. To do so, we plan to employ another algorithm that retrieves concepts that are related or co-exist with given concepts from a pre-trained deep language model.

\textit{Increasing the industry evaluation}. Several agile project management tools are used in the IT industry. To increase the industry adoption of \sys{}, we will create a way for integrating \sys{} with agile project management tools such as Jira.

\subsection{Threats to Validity}
This section covers the existing threats to \sys{} and, if possible, how those threats have been mitigated. We identify the threats based on the definitions of Wohlin \etal{} \cite{wohlin2012experimentation}.

\textit{Internal validity}. To prevent bias that may arise from using the publicly available user story sets, we design an experiment that requires participants to construct user stories while they are using different tools as \sys{} and Google Docs. Therefore, we are able to mimic a real use-case scenario. We also try to minimize the bias that might arise from the participant involvement. Participants are randomly assigned to different groups and decision for which tool to start is also randomly decided. Order of the available scenarios are also randomly assigned to groups. By doing so, we ensure that equal numbers of randomly assigned groups are introduced to \sys{} in different experiment sessions with different scenarios. Additionally, we choose the development of \textit{restaurant and library} management systems as our scenarios since they do not require any advanced level of expertise and these scenarios are widely used examples for educational purposes thus eliminating the risk of one group dominating another one.  When we consider all of the actions that we take, we apply these steps to eliminate any threats caused by learning effect.

We also consider the needs of participants while conducting the experiment. 
First, participants can easily reach our scenario when they are using Google Docs. By default, Google Docs is a text editor which allows them to read the scenario just by scrolling to the top without looking any other screen or window. To achieve the same accessibility level with \sys{}, we present the scenario on top of the user interface. 
Second, participants can easily communicate with each other while using Google Docs. Via communicating, participants can build consensus on some vague points such as deciding on an actor that represented by manager and restaurant manager by different stakeholders. To provide a communication channel to users, we implement a message broadcasting mechanism that allows them to communicate different stakeholders that work on the same project.

\textit{External validity}. In empirical software engineering research, it is crucial to demonstrate the generalizability of a study. We believe that our participant group is representative of the target user group for the experiment. We form a diverse participant group for the experiments. The participant group involves participants with different backgrounds and level of expertise. This may potentially reduce concerns about the validity of the study due to the small number of participants, as Falessi \etal{} \cite{falessi2018empirical} report that a small number of participants can produce credible results when they accurately represent the target user population. Yet, we seek support from the community so that different practitioners utilize \sys{} to prove its generalizability.

\textit{Construct validity}. In agile requirements engineering environments, collaborative writing editors are widely used to construct requirements artifacts. Google Docs provides favorable features for users to ease their work and to collaborate easily. We choose Google Docs which is a widely used collaborative writing editor as a benchmark for \sys{}. This decision is justifiable since Google Docs is a popular tool. Also, employing a generally accepted tool for our experiments reduces the threats that caused by participants that are unfamiliar with the collaborative writing editor. After completing our experiments, we choose statistical tests that comply with the obtained data. First, we check whether the samples belong to different tools come from a normal distribution. Second, we measure the statistical significance of our results by applying Mann-Whitney U Test. Since we collect two independent sets with relatively small size, Mann-Whitney U Test is the best fit for our data. We manage to achieve a p-value of 0.015 which is smaller than the alpha value of 0.05. This results indicates that \sys{} obtained a statistically different result compared to Google Docs.

\textit{Conclusion validity}. Although user studies are inherently subjective and cannot ensure the complete repeatability of results, we try to minimize this issue by sharing our results that are obtained after experiments. To guarantee the consistency of the \sys{}, we established a standardized experimental procedure during the study design and use it for every participant. We also publicly share different materials for replication such as user stories that are generated by participants, source code for the \sys{}.

\bibliographystyle{unsrt}
\bibliography{references}
\end{document}